\documentclass[aps,prd,superscriptaddress,twocolumn,preprintnumbers,floatfix]{revtex4-1}

\usepackage{graphicx}
\usepackage[small,bf]{subfigure}                           
\usepackage{dcolumn}                                       
\newcolumntype{x}{D{.}{.}{-1}}
\usepackage[thinspace,thinqspace,amssymb]{SIunits}         
\newcommand*{\flux}{\giga\electronvolt\reciprocal\steradian\reciprocal\second\centi\metre\rpsquared}

\makeatletter
\newcommand*{\rom}[1]{\expandafter\@slowromancap\romannumeral #1@}
\makeatother

\newcommand*{\Stephanie}{\rom{2}}
\newcommand*{\EikeLE}{\rom{1}a}
\newcommand*{\EikeHE}{\rom{1}b}
\newcommand*{\BDTI}{\mathrm{BDT}_\mathrm{II}}              
\newcommand*{\BDTII}{\mathrm{BDT}_\mathrm{I}}              
\newcommand*{\mcolon}{\mathbin{:}} 

\usepackage{etoolbox}                                      
\usepackage{amsmath}
\usepackage{amssymb}
\usepackage{multirow}
\apptocmd{\sloppy}{\hbadness 10000\relax}{}{}

\let\oldcite\cite
\renewcommand{\cite}{\!~\oldcite}                          

\usepackage[pdftex,
           raiselinks=true,%
           bookmarks=true,%
           bookmarksopenlevel=1,%
           bookmarksopen=true,%
           bookmarksnumbered=true,%
           hyperindex=true,%
           plainpages=false,
           pageanchor=true,
           linktocpage=true,
           pdfpagelabels,
           colorlinks=false,
           citecolor=black,
           urlcolor=black,
           hyperfootnotes=false]{hyperref} 

\begin{document}

\title{Search for neutrino-induced particle showers with IceCube-$\mathbf{40}$}

\affiliation{III. Physikalisches Institut, RWTH Aachen University, D-52056 Aachen, Germany}
\affiliation{School of Chemistry \& Physics, University of Adelaide, Adelaide SA, 5005 Australia}
\affiliation{Dept.~of Physics and Astronomy, University of Alaska Anchorage, 3211 Providence Dr., Anchorage, AK 99508, USA}
\affiliation{CTSPS, Clark-Atlanta University, Atlanta, GA 30314, USA}
\affiliation{School of Physics and Center for Relativistic Astrophysics, Georgia Institute of Technology, Atlanta, GA 30332, USA}
\affiliation{Dept.~of Physics, Southern University, Baton Rouge, LA 70813, USA}
\affiliation{Dept.~of Physics, University of California, Berkeley, CA 94720, USA}
\affiliation{Lawrence Berkeley National Laboratory, Berkeley, CA 94720, USA}
\affiliation{Institut f\"ur Physik, Humboldt-Universit\"at zu Berlin, D-12489 Berlin, Germany}
\affiliation{Fakult\"at f\"ur Physik \& Astronomie, Ruhr-Universit\"at Bochum, D-44780 Bochum, Germany}
\affiliation{Physikalisches Institut, Universit\"at Bonn, Nussallee 12, D-53115 Bonn, Germany}
\affiliation{Universit\'e Libre de Bruxelles, Science Faculty CP230, B-1050 Brussels, Belgium}
\affiliation{Vrije Universiteit Brussel, Dienst ELEM, B-1050 Brussels, Belgium}
\affiliation{Dept.~of Physics, Chiba University, Chiba 263-8522, Japan}
\affiliation{Dept.~of Physics and Astronomy, University of Canterbury, Private Bag 4800, Christchurch, New Zealand}
\affiliation{Dept.~of Physics, University of Maryland, College Park, MD 20742, USA}
\affiliation{Dept.~of Physics and Center for Cosmology and Astro-Particle Physics, Ohio State University, Columbus, OH 43210, USA}
\affiliation{Dept.~of Astronomy, Ohio State University, Columbus, OH 43210, USA}
\affiliation{Niels Bohr Institute, University of Copenhagen, DK-2100 Copenhagen, Denmark}
\affiliation{Dept.~of Physics, TU Dortmund University, D-44221 Dortmund, Germany}
\affiliation{Dept.~of Physics, University of Alberta, Edmonton, Alberta, Canada T6G 2E1}
\affiliation{Erlangen Centre for Astroparticle Physics, Friedrich-Alexander-Universit\"at Erlangen-N\"urnberg, D-91058 Erlangen, Germany}
\affiliation{D\'epartement de physique nucl\'eaire et corpusculaire, Universit\'e de Gen\`eve, CH-1211 Gen\`eve, Switzerland}
\affiliation{Dept.~of Physics and Astronomy, University of Gent, B-9000 Gent, Belgium}
\affiliation{Dept.~of Physics and Astronomy, University of California, Irvine, CA 92697, USA}
\affiliation{Laboratory for High Energy Physics, \'Ecole Polytechnique F\'ed\'erale, CH-1015 Lausanne, Switzerland}
\affiliation{Dept.~of Physics and Astronomy, University of Kansas, Lawrence, KS 66045, USA}
\affiliation{Dept.~of Astronomy, University of Wisconsin, Madison, WI 53706, USA}
\affiliation{Dept.~of Physics and Wisconsin IceCube Particle Astrophysics Center, University of Wisconsin, Madison, WI 53706, USA}
\affiliation{Institute of Physics, University of Mainz, Staudinger Weg 7, D-55099 Mainz, Germany}
\affiliation{Universit\'e de Mons, 7000 Mons, Belgium}
\affiliation{T.U. Munich, D-85748 Garching, Germany}
\affiliation{Bartol Research Institute and Dept.~of Physics and Astronomy, University of Delaware, Newark, DE 19716, USA}
\affiliation{Dept.~of Physics, University of Oxford, 1 Keble Road, Oxford OX1 3NP, UK}
\affiliation{Dept.~of Physics, University of Wisconsin, River Falls, WI 54022, USA}
\affiliation{Oskar Klein Centre and Dept.~of Physics, Stockholm University, SE-10691 Stockholm, Sweden}
\affiliation{Dept.~of Physics and Astronomy, Stony Brook University, Stony Brook, NY 11794-3800, USA}
\affiliation{Dept.~of Physics, Sungkyunkwan University, Suwon 440-746, Korea}
\affiliation{Dept.~of Physics, University of Toronto, Toronto, Ontario, Canada, M5S 1A7}
\affiliation{Dept.~of Physics and Astronomy, University of Alabama, Tuscaloosa, AL 35487, USA}
\affiliation{Dept.~of Astronomy and Astrophysics, Pennsylvania State University, University Park, PA 16802, USA}
\affiliation{Dept.~of Physics, Pennsylvania State University, University Park, PA 16802, USA}
\affiliation{Dept.~of Physics and Astronomy, Uppsala University, Box 516, S-75120 Uppsala, Sweden}
\affiliation{Dept.~of Physics, University of Wuppertal, D-42119 Wuppertal, Germany}
\affiliation{DESY, D-15735 Zeuthen, Germany}

\author{M.~G.~Aartsen}
\affiliation{School of Chemistry \& Physics, University of Adelaide, Adelaide SA, 5005 Australia}
\author{R.~Abbasi}
\affiliation{Dept.~of Physics and Wisconsin IceCube Particle Astrophysics Center, University of Wisconsin, Madison, WI 53706, USA}
\author{M.~Ackermann}
\affiliation{DESY, D-15735 Zeuthen, Germany}
\author{J.~Adams}
\affiliation{Dept.~of Physics and Astronomy, University of Canterbury, Private Bag 4800, Christchurch, New Zealand}
\author{J.~A.~Aguilar}
\affiliation{D\'epartement de physique nucl\'eaire et corpusculaire, Universit\'e de Gen\`eve, CH-1211 Gen\`eve, Switzerland}
\author{M.~Ahlers}
\affiliation{Dept.~of Physics and Wisconsin IceCube Particle Astrophysics Center, University of Wisconsin, Madison, WI 53706, USA}
\author{D.~Altmann}
\affiliation{Erlangen Centre for Astroparticle Physics, Friedrich-Alexander-Universit\"at Erlangen-N\"urnberg, D-91058 Erlangen, Germany}
\author{C.~Arguelles}
\affiliation{Dept.~of Physics and Wisconsin IceCube Particle Astrophysics Center, University of Wisconsin, Madison, WI 53706, USA}
\author{T.~C.~Arlen}
\affiliation{Dept.~of Physics, Pennsylvania State University, University Park, PA 16802, USA}
\author{J.~Auffenberg}
\affiliation{Dept.~of Physics and Wisconsin IceCube Particle Astrophysics Center, University of Wisconsin, Madison, WI 53706, USA}
% ...

\date{\today}

\collaboration{IceCube Collaboration}
\noaffiliation

\date{\today}

\begin{abstract}
We report on the search for neutrino-induced particle-showers, so-called
cascades, in the IceCube-$40$ detector. The data for this search was collected
between April 2008 and May 2009 when the first $40$ IceCube strings were
deployed and operational. Three complementary searches were performed, each
optimized for different energy regimes. The analysis with the lowest energy
threshold ($\unit{2}{\tera\electronvolt}$) targeted atmospheric neutrinos. A
total of $67$ events were found, consistent with the expectation of $41$
atmospheric muons and $30$ atmospheric neutrino events. The two other analyses
targeted a harder, astrophysical neutrino flux. The analysis with an
intermediate threshold of $\unit{25}{\tera\electronvolt}$ lead to the
observation of $14$ cascade-like events, again
consistent with the prediction of $3.0$ atmospheric neutrino and $7.7$
atmospheric muon events. We hence set an upper limit of $\unit{E^2 \Phi_{lim}
\leq 7.46\times10^{-8}}{\flux}$ ($90\%$ C.L.) on the diffuse flux from
astrophysical neutrinos of all neutrino flavors, applicable to the energy range
$\unit{25}{\tera\electronvolt}$ to $\unit{5}{\peta\electronvolt}$, assuming an
$E_{\nu}^{-2}$ spectrum and a neutrino flavor ratio of $1\mcolon1\mcolon1$ at
the Earth. The third analysis utilized a larger and optimized sample of
atmospheric muon background simulation, leading to a higher energy threshold
of $\unit{100}{\tera\electronvolt}$. Three events were found over a background
prediction of $0.04$ atmospheric muon events and $0.21$ events from the flux of
conventional and prompt atmospheric neutrinos. Including systematic errors this
corresponds to a $\unit{2.7}{\sigma}$ excess with respect to the background-only
hypothesis. Our observation of neutrino event candidates above
$\unit{100}{\tera\electronvolt}$ complements IceCube's recently observed
evidence for high-energy astrophysical neutrinos.
\end{abstract}

\maketitle


\section{Introduction} \label{Section:Introduction}
One century after the discovery of cosmic rays the search for their sources is
still ongoing. Astrophysical objects which are either confirmed or expected to
be able to accelerate hadrons to the observed energies include supernova
remnants\cite{Ackermann:2013wqa}, active galactic nuclei (AGN), gamma-ray bursts
(GRBs), and shocks in star formation regions of galaxies. The cosmic-ray nuclei
interact with ambient matter and radiation fields close to their
source\cite{Hillas:2006ms}. Charged pions produced in these interactions decay
into neutrinos. Therefore, the detection of high-energy neutrinos from such
objects provides a unique possibility to identify individual astrophysical
objects as cosmic-ray sources. However, low fluxes and small interaction
probabilities make the detection of high-energy neutrinos challenging. To date
no astrophysical object has been conclusively identified as a source of
$\unit{}{\tera\electronvolt}$ neutrinos. Previous searches have established
limits enabling astrophysical models to be
constrained\cite{Abbasi:2012zw,pointsource}. 

While individual neutrino sources might be too weak to be detectable with
current instruments, they would still contribute to a collective astrophysical
neutrino flux. Fermi shock acceleration is thought to be the main acceleration
mechanism for cosmic-ray nuclei and therefore a power-law spectrum with an index
of about $-2$ is expected for the nuclei in the interaction regions where the
neutrinos are produced. Based on the energy density of ultra-high-energy cosmic
rays and assuming the cosmic-ray sources are transparent, the all-flavor diffuse
neutrino flux can be constrained theoretically to be lower than the
Waxman-Bahcall bound of $\unit{E_\nu^{2}\Phi \lessapprox
3\times10^{-8}}{\flux}$\cite{Waxman:1998yy,Waxman:2011hr}. As neutrinos are
assumed to originate mainly from pion decays, at the source a flavor ratio of
$\nu_{e}\mcolon\nu_{\mu}\mcolon\nu_{\tau} = 1\mcolon2\mcolon0$ is expected. This
ratio would transform to $1\mcolon1\mcolon1$ on Earth due to neutrino
oscillations\cite{Athar:2000yw,Beacom:2003nh}. However, observing unequal or
energy-dependent flavor contributions would be interesting, since for example
the flavor ratio is sensitive to the assumed production mechanism at the
source\cite{Kashti:2005qa}.

\begin{figure}[tb]
\includegraphics[width=0.49\textwidth]{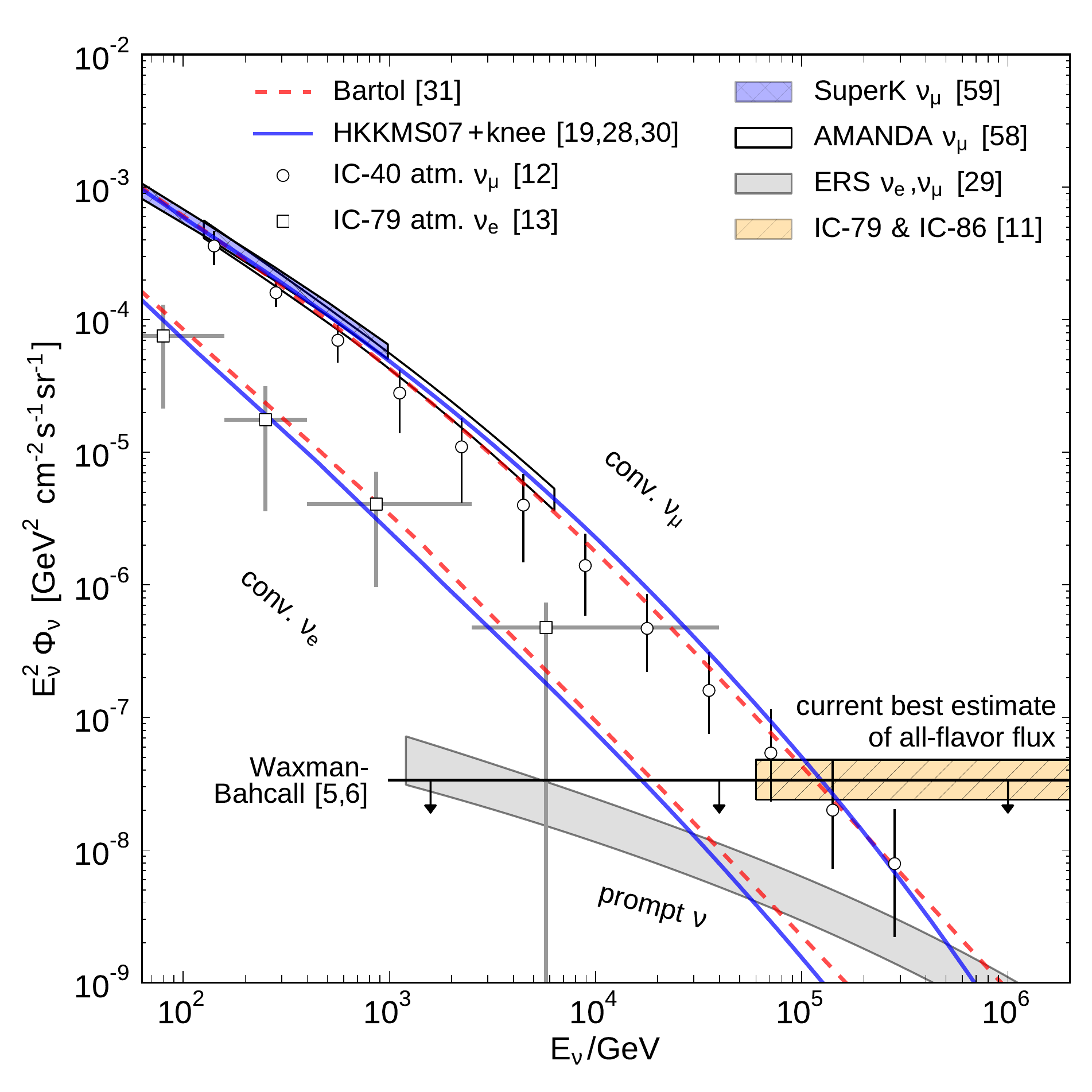}
\caption{Neutrino energy spectrum above $\unit{100}{\giga\electronvolt}$. The
theoretical predictions and measurements for the atmospheric neutrino flux are
shown, as well as the current estimate for the diffuse astrophysical neutrino flux.}
\label{Figure:AtmosphericDiffuseFluxes}
\end{figure}

Recently, evidence for this diffuse astrophysical neutrino flux was found. Its
all-flavor intensity is estimated to be $\unit{E_\nu^{2}\Phi =
(3.6\pm1.2)\times10^{-8}}{\flux}$ with indications for a cutoff at
$\unit{\sim2}{\peta\electronvolt}$. It is consistent with an isotropic flux and
a flavor ratio of $1\mcolon1\mcolon1$ (\cite{Aartsen:2013bka,hese} and
Fig.~\ref{Figure:AtmosphericDiffuseFluxes}).

In order to measure the diffuse astrophysical neutrino flux at
$\tera\electronvolt$ energies, it has to be separated from two main sources of
background, which both originate from the Earth's atmosphere. These are
atmospheric muons and neutrinos produced in cosmic-ray air showers. The
atmospheric neutrino flux has two components. The so-called conventional
atmospheric neutrinos are produced in decays of pions and kaons. Their intensity
is well-measured up to $\unit{6}{\tera\electronvolt}$ for $\nu_{e}$ and up to
$\unit{400}{\tera\electronvolt}$ for
$\nu_{\mu}$\cite{Abbasi:2010ie,Aartsen:2012uu}. At higher energies the poor
knowledge of the composition of the cosmic-ray flux, creating the neutrinos,
causes significant uncertainties on the intensity. The spectrum of the
conventional atmospheric neutrinos is steeper than the cosmic-ray spectrum due
to pion and kaon energy losses in the atmosphere. The second component
originates from the decay of charmed mesons, which have livetimes several orders
of magnitude smaller than charged pions and kaons. Accordingly, neutrinos from
these decays are called prompt atmospheric neutrinos. Due to the short lifetime
of the parent mesons the energy spectrum of the prompt atmospheric neutrinos is
expected to follow the spectrum of the cosmic rays that create them. However,
their intensity has never been measured and uncertainties in the relevant
production cross sections lead to large uncertainties in the predicted flux. The
presence of a prompt neutrino component, like an astrophysical neutrino
component, introduces a break into the neutrino energy spectrum. Given the
large uncertainties in the prompt neutrino predictions, identification and
separation of the astrophysical and prompt components needs to be made through
their respective spectral signatures (see Fig.~\ref{Figure:AtmosphericDiffuseFluxes}).

The IceCube Neutrino Observatory is located at the South Pole and is the first
kilometer-scale Cherenkov neutrino telescope. An optical sensor array observes
the Cherenkov radiation from secondary charged particles produced in neutrino
interactions deep in the ice. These are dominantly neutrino-nucleon
interactions except for the Glashow resonance\cite{Glashow:1960} for electron
anti-neutrinos at $\unit{6.3}{\peta\electronvolt}$. Based on the signature of
the neutrino interaction, which depends on the flavor of the incident neutrino
and the type of the interaction, two main detection channels exist. Searches in
the muon channel look for charged-current muon neutrino interactions. These have
a muon in the final state whose direction is reconstructible with a resolution
of about $1$ degree\cite{Aartsen:2013zka}. The large muon range also allows
to detect neutrino interactions outside the instrumented volume. The cascade
channel comprises all other interaction scenarios which have particle showers
in the final state. Above \unit{}{\peta\electronvolt} energies charged current 
$\nu_\tau$ interactions exhibit more complex event signatures but at lower 
energies they appear in the cascade channel. Consequently, an astrophysical flux
with equal neutrino flavor contributions would yield more cascade than track
events starting inside of IceCube.
If the neutrino interaction happens inside the detector, the Cherenkov light
yield of particle showers scales nearly linearly with the deposited energy,
leading to an energy resolution that is better than in the muon channel. On the
other hand the angular resolution is rather poor ($>\unit{10}{\degree}$ for the
completed IceCube detector). Overall, the cascade channel is best suited for
searches for diffuse astrophysical neutrinos in which the neutrino energy
measurement is more important than pointing capabilities\cite{Kowalski:2005tz}.

\nocite{Abbasi:2011jx} 

This paper presents searches for neutrino-induced cascades in one year of data
taken during the construction phase of IceCube, when about half the detector
was operational (IceCube-$40$). The main objective of the searches was to
identify an astrophysical flux of neutrinos. In addition, a sensitivity to
atmospheric neutrinos in the few $\unit{}{\tera\electronvolt}$ energy range was
maintained to allow a validation of the anticipated backgrounds in the data set.

A $2.7\sigma$ excess of events above $\unit{100}{\tera\electronvolt}$ was
found, compatible with the all flavor astrophysical diffuse neutrino flux
estimate obtained in IceCube's high-energy starting events (HESE)
analysis\cite{hese}. In comparison to that analysis, the IceCube-$40$ cascade
analysis provides an event sample with unprecedented low background
contamination between $100$ and $\unit{200}{\tera\electronvolt}$. This is
possible because both searches employed rather different event selection
strategies. Methods outlined in this paper also prove powerful in cascade
searches with later IceCube
configurations\cite{Schoenwald:2013ab,Bzdak:2013ab}.

The paper is organized as follows: The IceCube detector and IceCube-$40$
dataset are described in section \ref{Section:TheIceCubeDetector}. The
simulation used is presented in section \ref{Section:Simulation} followed by a
description of the cascade reconstruction in section
\ref{Section:ReconstructionOfParticleShowers}. The details of the event
selection and expected sensitivity are presented in section
\ref{Section:EventSelectionAndAnalysisMethod}. A survey of the systematic
uncertainties follows in section \ref{Section:SystematicUncertainties} before
the results and implications are discussed in section
\ref{Section:ResultsAndDiscussion}. A conclusion is given in section
\ref{Section:Summary}.


\section{The IceCube Detector} \label{Section:TheIceCubeDetector}

\begin{figure}[tb]
\includegraphics[width=0.49\textwidth]{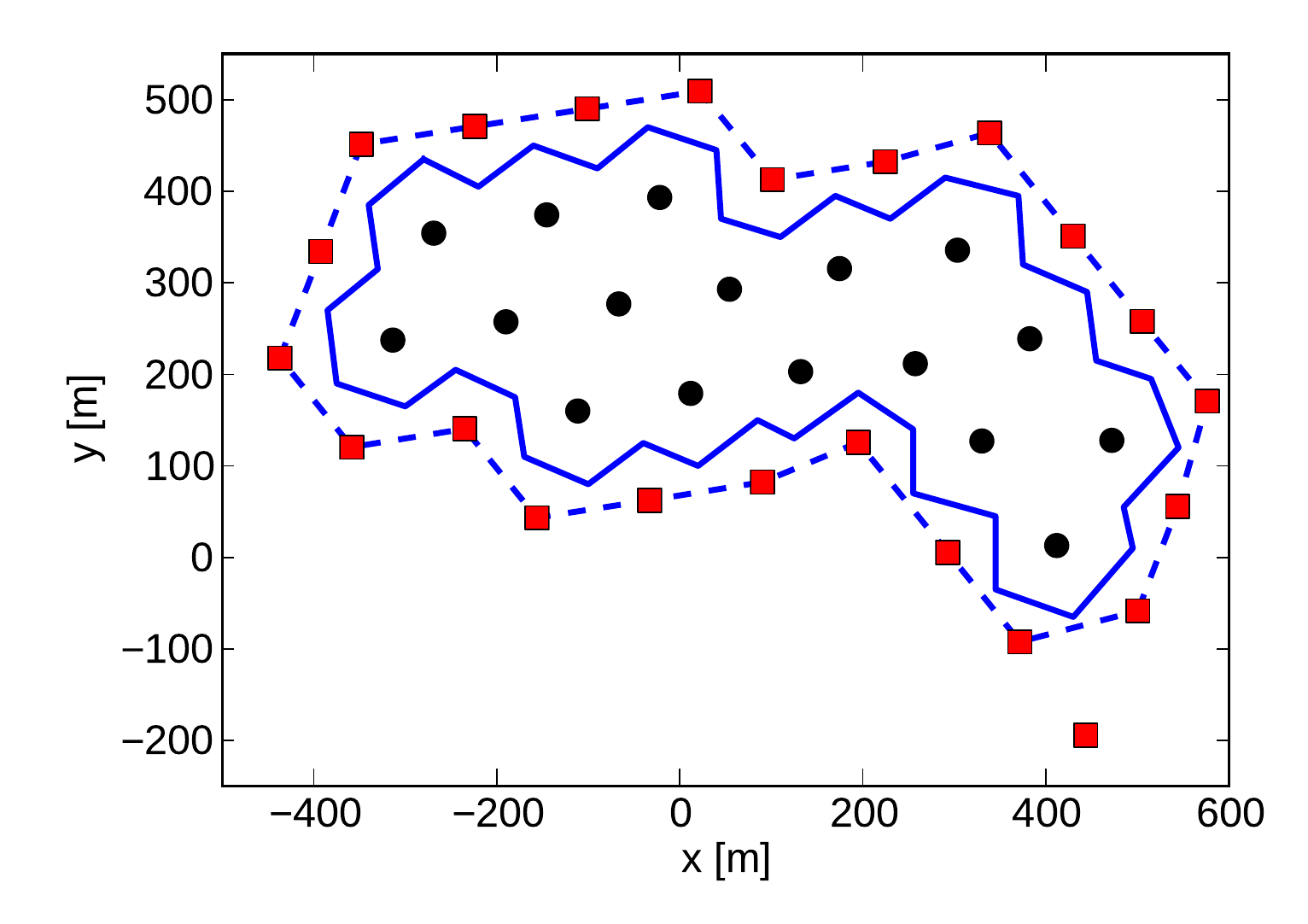}
\caption{The IceCube-$40$ detector configuration. The circles and squares are
the positions of the strings. The point $(x,y) = (0,0)$ is the centre of the
complete $86$-string detector. Particle showers with reconstructed vertices
inside the instrumented volume are called contained events. The analyses
presented in this work reject non-contained events in order to suppress
atmospheric muons that enter the detector from the outside. The blue dashed and
solid lines show the two differently tight containment requirements that are
used. The strings denoted by red squares form the outer layer of the detector.
They are used to veto incident atmospheric muons.}
\label{Figure:IC40detector}
\end{figure}
The IceCube Neutrino Observatory\cite{Achterberg2006155} consists of an in-ice
array of optical sensors and a complementary surface air shower detector called
IceTop. The analyses presented here utilized only the in-ice component so the
following detector description will be limited to that.

The optical sensors, called Digital Optical Modules (DOMs)\cite{Abbasi:2008aa},
are sensitive to Cherenkov photons between $\unit{350}{\nano\meter}$ and
$\unit{650}{\nano\meter}$. The DOMs are deployed between depths of
$\unit{1450}{\metre}$ and $\unit{2450}{\metre}$ and are attached to strings that
are formed by the readout cables. Each string has $60$ DOMs attached. 
The vertical string spacing of the DOMs is approximately $\unit{17}{\metre}$
and the horizontal spacing between the strings is approximately
$\unit{125}{\metre}$. The data for this analysis was collected between April
2008 and May 2009 with a total of $367.1$ days livetime. In this period 
$40$ strings were deployed and operational. The detector layout is shown in
Fig.~\ref{Figure:IC40detector}. Before IceCube's completion in 2010, the
IceCube-$59$, and IceCube-$79$ configurations took data with $59$ and $79$
deployed strings, respectively.

Each DOM consists of a $\unit{25}{\centi\metre}$ diameter photomultiplier Tube 
(PMT)\cite{Abbasi:2010vc}, made by Hamamatsu Photonics, and a data acquisition
system housed within a pressure sphere made of $\unit{13}{\milli\metre}$ thick
borosilicate glass. The PMT's dynamic range is $200$ photoelectrons per
$\unit{15}{\nano\second}$ and it is designed to accurately record the amplitudes
and widths of the pulses with a timing resolution of $\unit{5}{\nano\second}$.
Their peak quantum efficiency is approximately $25\%$ and they are
operated at a gain of $10^{7}$ to resolve single photoelectrons.

The time-resolved PMT signal (waveform) is digitized in the DOM. For this
purpose two digitization devices are available on the DOM mainboard: two Analog
Transient Waveform Digitizers (ATWD) and a fast Analog-to-Digital Converter 
(fADC). The ATWDs have three channels operated in parallel at different gains to
provide a large dynamic range (a fourth channel is used only for calibration
purposes). Due to scattering in the ice the arrival times of photons emitted at
the same point and time can vary by microseconds. The ATWDs provide a sampling
rate of $300\,\textrm{Megasamples/s}$ over a time window of
$\unit{425}{\nano\second}$ allowing them to record the earliest photons (i.e.
those least affected by scattering in the ice) with high precision. The second
digitizer, the fADC, has a coarser sampling of $40\,\textrm{Megasamples/s}$
recording data over a longer time period for photons with larger delays of up to
$\unit{6.4}{\micro\second}$. In order to reduce data readout volume due to
noise, in IceCube-$40$ a local coincidence criterion is required. Only if a
neighbouring DOM on the same string also detects light within the local
coincidence time window of $\pm\unit{1000}{\nano\second}$, the PMT response is
digitized, time-stamped, and transmitted to the surface for analysis. The
surface data acquisition system combines the individual PMT responses and forms
events when one of the several possible triggering criteria are fulfilled.

The trigger requirement for the IceCube-$40$ cascade search was the Simple
Multiplicity Trigger, which requires that eight DOMs were hit within a
$\unit{5000}{\nano\second}$ time window. The data rate for IceCube-$40$ from
this trigger was approximately $\unit{1000}{\hertz}$. 


\section{Simulation} \label{Section:Simulation}

Interactions of all flavors of neutrinos were simulated to model atmospheric and
astrophysical neutrinos. The \texttt{NuGEN} software package maintained by the
IceCube collaboration was used. It is based on the
\texttt{ANIS}\cite{Gazizov:2004va} neutrino generator, which produces neutrinos
isotropically over the Earth's surface and propagates them to interact in or
near the detector volume. Neutrino attenuation and $\nu_{\tau}$ regeneration are
accounted for using the PREM Earth model\cite{Dziewonski:1981xy}. CTEQ5
structure functions\cite{Lai:1999wy} were used to model the deep-inelastic
neutrino-nucleon scattering cross section. 

Throughout this paper the diffuse astrophysical neutrino flux is simulated
isotropically, with a flavor ratio of $1\mcolon1\mcolon1$ and, if not stated
otherwise, with an unbroken power-law spectrum with index of $-2$ and an all-flavor
intensity of $\unit{3.6\times10^{-8}}{\flux}$.

Rate predictions for the atmospheric neutrinos are based on the HKKMS07 
model\cite{Honda:2006qj} for conventional atmospheric neutrinos
and the ERS model\cite{Enberg:2008te} for prompt atmospheric
neutrinos. Extrapolations of the original calculations to higher energies
provide rate predictions at the energy range relevant to this work. The
steepening of the cosmic-ray spectrum around a few $\unit{}{\peta\electronvolt}$
(the so-called ``knee'') causes a similar feature in the atmospheric neutrino
spectrum which is not accounted for in the HKKMS07 model. A modification to the
HKKMS07 model\cite{Gaisser:2012zz,Schukraft:2013ya} was applied to account for
the knee. For one of the presented analyses the Bartol
model\cite{Barr:2004br} was used to estimate the conventional atmospheric
neutrino flux. Compared to the modified HKKMS07 model it predicts a higher
$\nu_{e}$ contribution (see Fig.~\ref{Figure:AtmosphericDiffuseFluxes}).

The propagation of muons and taus through the detector and their energy losses
were simulated using the \texttt{MMC} program\cite{Chirkin:2002xf} and the
cascade simulation inside the detector was handled by the \texttt{CMC}
program\cite{Voigt:2008zz}. Neutrino-induced cascades below a threshold of
$\unit{1}{\tera\electronvolt}$ were simulated as point-like light sources,
emitting an angular Cherenkov light profile typical of an electromagnetic
shower\cite{Radel:2012ij}. Cascades of higher energies are split into segments
along the direction of the shower development. Each cascade segment is then
approximated by a point-like sub-shower with a light yield proportional to the
light yield in the corresponding segment of the electromagnetic cascade.
The elongation of electromagnetic cascades due to the suppression of 
bremsstrahlung and pair production cross sections above
$\unit{}{\peta\electronvolt}$ energies (LPM effect\cite{Gerhardt:2010bj}) is
accounted for. Hadronic cascades are simulated as electromagnetic cascades with
a smaller light yield per deposited energy to account for the neutral shower
components which do not generate Cherenkov light\cite{Kowalski:2002aa}.

The contribution from atmospheric muon events is estimated from simulations done
with a modified version\cite{Chirkin:2004ic,Chirkin:2003qn} of the
\texttt{CORSIKA} air shower simulation software\cite{Heck:1998vt}. A large
number of background events must be generated due to the high background
suppression that is necessary to reach an event sample dominated by neutrinos.
Providing a large background sample is computationally challenging, mostly
because of the sheer number of air showers needed but also due to the simulation
of light propagation in the optically inhomogeneous ice. The figure of merit
used to quantify the statistics of a simulated data sample is the effective
livetime $T_{\mathrm{eff}}$, i.e. the time that one would have to run the real
experiment to obtain the same statistical error as in the simulated dataset. 

\begin{figure}[tb]
\includegraphics[width=0.49\textwidth]{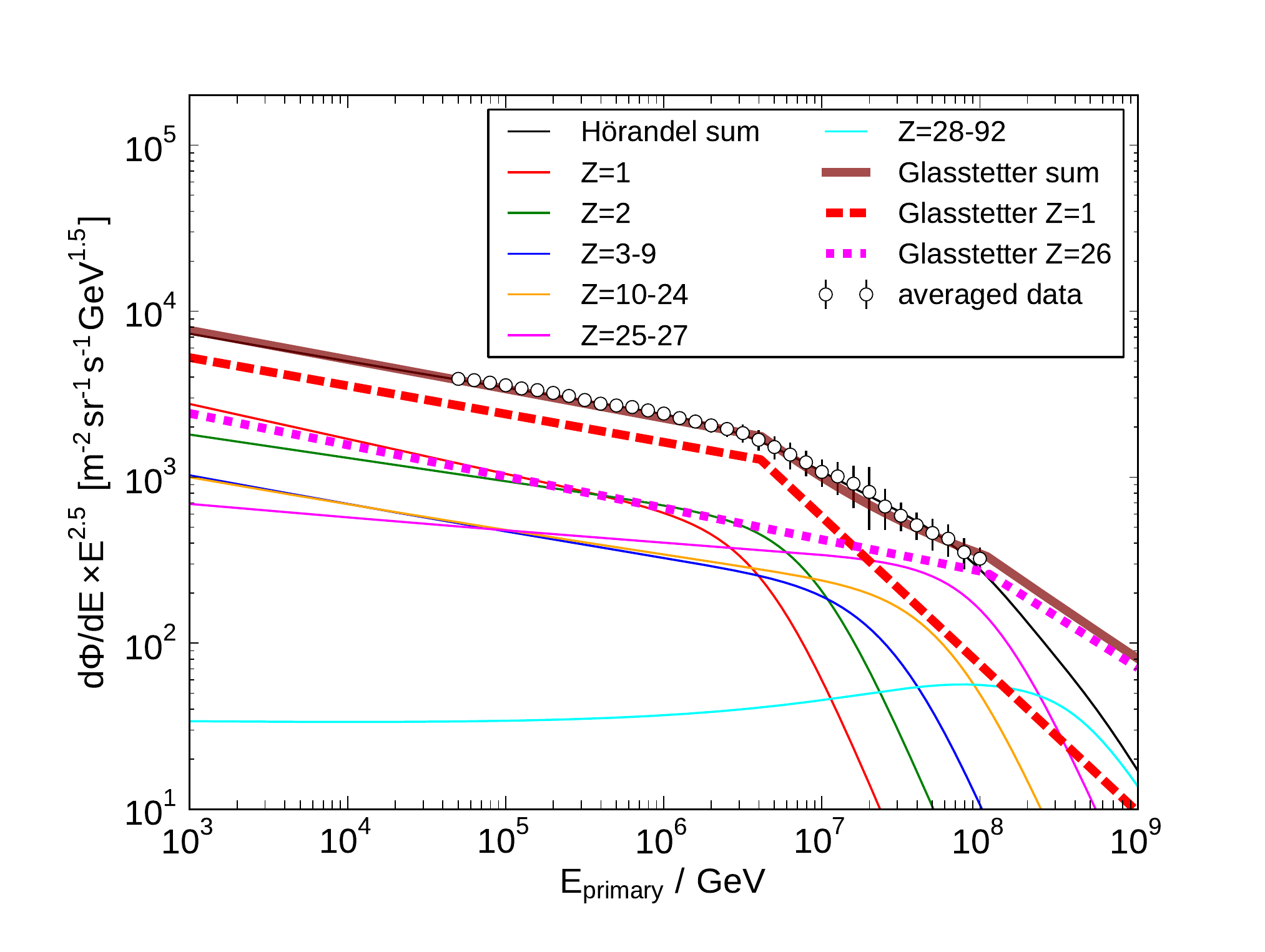}
\caption{The cosmic-ray spectrum is modeled with two broken power laws for
proton and iron (red and magenta dashed lines). Their parameters are taken
from\cite{Glasstetter:1999ua}. The H\"orandel model and the data from which it
has been derived is shown for comparison (both taken
from\cite{Hoerandel:2002yg}). The H\"orandel model comprises components for each
element from hydrogen up to iron illustrated by solid lines.}
\label{Figure:EikeCRSpectrum}
\end{figure}

\begin{figure}[tb]
\includegraphics[width=0.49\textwidth]{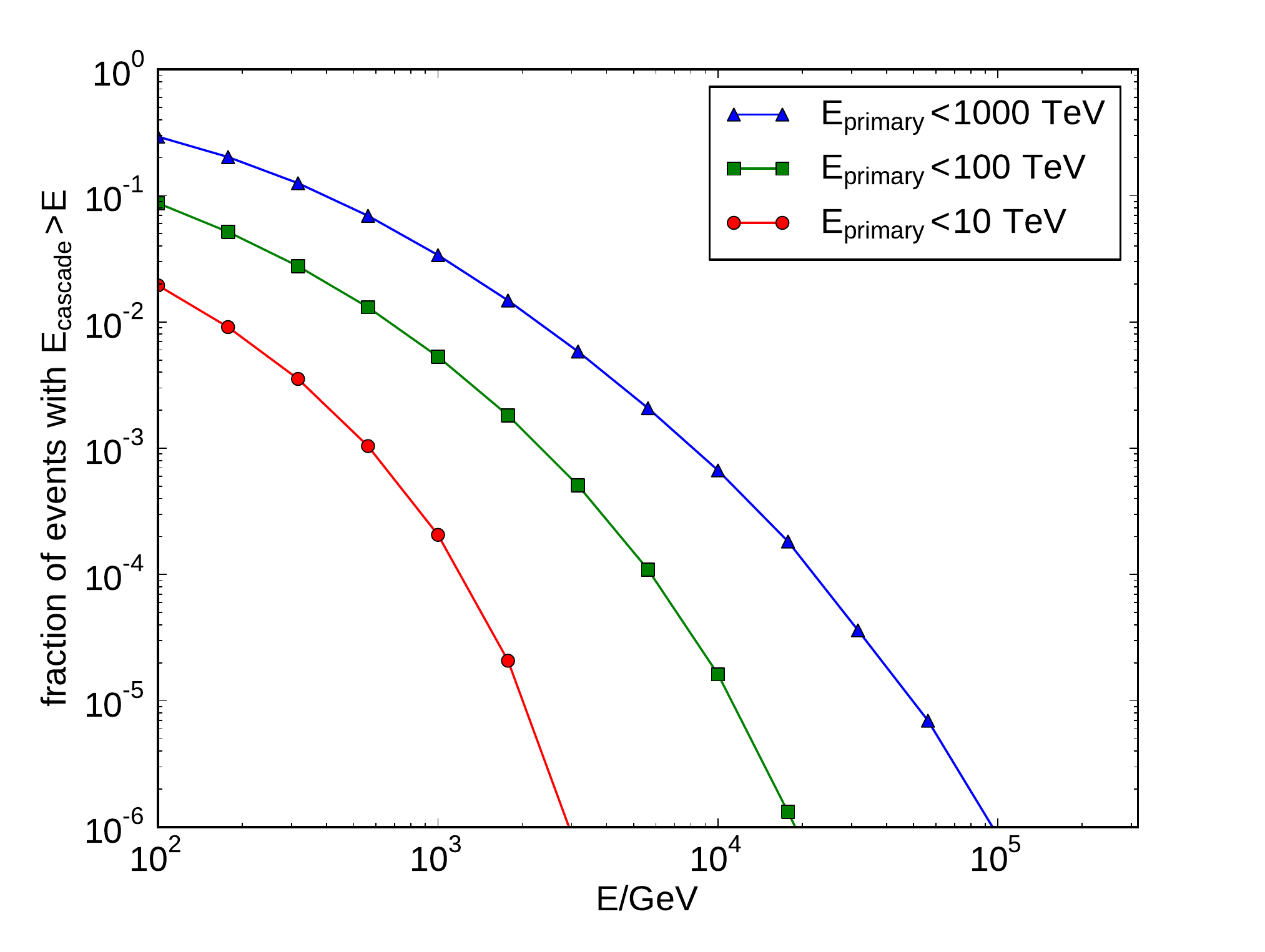}
\caption{Proton air showers are simulated according to the Glasstetter spectrum
illustrated in Fig.~\ref{Figure:EikeCRSpectrum}. Bremsstrahlung cascades from
muons originating in these showers are studied in order to assess the
probability to obtain a bremsstrahlung cascade bright enough to pass event
selection cuts from air showers below a given energy threshold. The plot shows
the fraction of proton air showers with primary energy $E_\textrm{primary}$
below $10$, $100$, $\unit{1000}{\tera\electronvolt}$ that exhibit catastrophic
energy losses of $E_\textrm{cascade}$ above a given energy.}
\label{Figure:ecascfraction}
\end{figure}

The chemical composition of the cosmic rays is important for an estimate of the
muon background. Previous cascade searches\cite{Abbasi:2011ui} have shown that
protons are of prime importance. Proton primaries produce lower multiplicities
of muons with higher individual energies than heavier primaries. Compared to air
showers induced by heavier elements, which typically lead to muon bundles
reaching the detector, proton showers are more likely to generate single
high-energy muons. If such a high-energy muon has a single catastrophic
energy loss, then the relative light yield between shower and muon can make the
resulting event signature look shower-like. 

We use a simplified cosmic-ray composition model in our simulation that
nevertheless reflects these qualitative differences between light and heavy
elements. This two-component model considers only the two extrema of cosmic-ray
composition and comprises two broken power laws, one for the proton and one for
the iron spectrum, respectively (see\cite{Glasstetter:1999ua} and
Fig.~\ref{Figure:EikeCRSpectrum}). The mean logarithmic mass of the all-particle
spectrum formed from the sum of the two power laws is compatible with
measurements of this quantity by air shower experiments\cite{Hoerandel:2002yg}.

The created background sample provides more than a year of effective livetime
above an energy per nucleon threshold of about $E/A >
\unit{90}{\tera\electronvolt}$. Below this threshold the statistics drop rather
quickly and these low energetic events are underrepresented in the event sample.
However, the risk of underestimating the background is low, since the fraction
of air showers with a primary energy below $\unit{90}{\tera\electronvolt}$ which
exhibit bremsstrahlung losses of $\unit{2}{\tera\electronvolt}$
($\unit{10}{\tera\electronvolt}$) is only $1.8\times10^{-3}$
($1.6\times10^{-5}$) (see Fig.~\ref{Figure:ecascfraction}).


\section{Reconstruction of Particle Showers} \label{Section:ReconstructionOfParticleShowers}
Cherenkov light, emitted by charged particles, traverses the optically
inhomogeneous detector material and is then sampled by the three-dimensional
IceCube array with nanosecond precision. Hence, the light's arrival times and
amplitudes are the sole available information on any particle interaction in the
detector. Previous analyses relied on reconstruction algorithms that used only
parts of this information, i.e. by considering only the integrated amplitude of
all light reaching a DOM or by using only the arrival time of the first photon. 
A new algorithm \texttt{CREDO} was developed. By using the waveform information
\texttt{CREDO} is able to reconstruct not only the interaction vertex and the
deposited energy but also the direction of the incident neutrino. \texttt{CREDO}
is the first IceCube cascade reconstruction algorithm which is able to estimate
the direction of the neutrino. The measurement process is described with a
probabilistic model and a maximum likelihood estimator allows information on the
vertex of the neutrino interaction, the neutrino direction and the deposited
energy to be inferred.

\texttt{CREDO} considers the information on an event in the form of time
intervals $(t, t+\Delta t)$ in which a DOM at position $\vec{x}$ recorded a
given amount of charge $n$. Formally, the event is then described with the set
$\lbrace n_{i}(\vec{x},t,\Delta t)\rbrace$, where the index $i$ runs over all
such intervals. It should be noted, that this set also contains time intervals
in which no charge has been recorded, since these time intervals also carry
constraining information. The particle showers are modeled as point-like
Cherenkov emitters, in the same way as they are described in simulation. A
particle shower is then fully specified by $7$ parameters
$\alpha=(t,x,y,z,\theta,\phi,E)$: the time and position of the vertex, two
angles for the direction of the neutrino and the deposited energy. The
scattering and absorption of light in the Antarctic ice are depth dependent. The
\texttt{PHOTONICS} package\cite{Lundberg:2007mf} is used to calculate the light
propagation in the ice and to create tables with light arrival times and
amplitudes as a function of the depth and the relative position between particle
shower and DOM. In order to avoid binning effects the table values are
interpolated by multidimensional spline fits\cite{Whitehorn:2013aa} prior to
being used in the reconstruction. 

The \texttt{PHOTONICS} tables are used to calculate for each time interval $i$
an estimate of the mean expected charge $\mu_{i} = \hat{n}_{i}(\vec{x}, t,
\Delta t, \alpha) + r_{noise}\Delta t$~. In this sum $\hat{n}_{i}$ is the
contribution due to the cascade $\alpha$ and $r_{noise}\Delta t$ is a
continuous noise contribution. In each time interval a counting experiment is
performed, and the probability of a given measurement $\lbrace n_{i} \rbrace$
under the condition $\alpha$ can be calculated:
\begin{equation}
P(\lbrace n_{i} \rbrace | \alpha) = \prod_{i} \frac{\mu_{i}(\alpha)^{n_{i}}}{n_{i}!}\exp\left( -\mu_{i}(\alpha)\right) .
\end{equation}
From this probability one can construct a maximum likelihood estimator, yielding
the parameters $\hat{\alpha}$ best supported by the measurement:
\begin{equation}
\hat{\alpha} = \underset{\alpha}{\mathrm{argmin}}\,\, \mathcal{L}(\alpha) \quad\mathrm{where}\quad \mathcal{L}(\alpha) =-\log P(\lbrace n_{i} \rbrace | \alpha).
\end{equation}

The search for the global maximum in the $7$-dimensional likelihood space is
performed by minimizing the negative logarithm of the likelihood using the
SIMPLEX minimizer in the Minuit software package\cite{James:1975dr}. In order
to avoid local minima the search is done iteratively, where in each step the
minimizer starts at a different position in the parameter space. The iterative
minimization and the many necessary table lookups for each time interval form a
time-consuming procedure that cannot be applied to all events. However, this
reconstruction finds the position of the particle shower with a resolution of
$\unit{15}{\metre}$ horizontally and, due to the smaller DOM spacing,
$\unit{5}{\metre}$ vertically. The energy resolution for an astrophysical
neutrino spectrum is about $40\%$ and the angular resolution is about
$\unit{30}{\degree}$. 

Other variants of the likelihood reconstruction described here have been
developed and been used in more recent analyses. By incorporating improvements
in the understanding of the detector response and a better modeling of the ice
they provide better resolutions and determine IceCube's performance on cascades
today\cite{energyrecopaper}.


\section{Event Selection \& Analysis Method} \label{Section:EventSelectionAndAnalysisMethod}
A small number of neutrino generated showers need to be isolated from a large
background of atmospheric muons. This is achieved by finding and applying a set
of conditions on the reconstructed event properties that neutrino induced
showers fulfill, but atmospheric muons do not. In practice this selection is
implemented as a multi-step process. The first steps in the process are aimed
at conservatively reducing the background, allowing the use of computationally
intensive reconstruction algorithms, that can be applied to a reduced set of
events only. Each step in the event selection process is referred to as a
``Level''. Level $1$, the trigger condition, was described in section
\ref{Section:TheIceCubeDetector}. Level $2$, the online filter applied at South
Pole, and subsequent levels are outlined in this section. The section will
first introduce the classification schemes and cut variables used for
separation of the signal from the background and then turn to a description of
the individual samples.


\subsection{Background rejection methods and event property variables} \label{Section:Variables}
The background rejection criteria used in neutrino cascade searches can be
classified conceptually as belonging in the following four categories.

{\bfseries Reconstructed particle parameters:}
Four different likelihood reconstruction routines are used in the searches
described here. The routines differ in runtime, precision, initial assumptions
and in the number of neutrino parameters that they infer from the event. Three
of the reconstruction routines return parameters with the hypothesis that the
event is a particle-shower. The fourth routine used assumes the event contains
a muon track.

At earlier cut levels the \texttt{cscd\_llh} likelihood
reconstruction\cite{Abbasi:2011zzy} is used. It doesn't account for the optical
inhomogeneities of the ice but provides a quick estimate for the vertex and a
quality parameter $\textrm{rlogL}_\textrm{cscd}$ that describes how well the
event fits to the cascade hypothesis. 

Using the vertex estimate from \texttt{cscd\_llh} another estimate for the
deposited energy, $E_\textrm{ACER}$, is provided by the
\texttt{AtmCscdEnergyReco} algorithm\cite{D'Agostino:2009tf}. It is quick to
compute and considers the optical inhomogeneities of the ice.

At later levels the \texttt{CREDO} algorithm described in section
\ref{Section:ReconstructionOfParticleShowers} provides the best estimates for
the reconstructed vertex $\vec{x}_\textrm{CREDO}^{(n)}$, deposited energy
$E_\textrm{CREDO}^{(n)}$ and zenith angle $\Theta_\textrm{CREDO}^{(n)}$. The
superscript $(n)$ distinguishes, where necessary, applications of the
\texttt{CREDO} algorithm with differing numbers of iterations.

All events are also reconstructed with another likelihood
reconstruction\cite{Ahrens:2003fg} which assumes that the event contains a muon
track. This routine gives a zenith angle estimate $\Theta_\textrm{track}$ and
provides a quality of fit parameter $\textrm{rlogL}_\textrm{track}$ for the
track hypothesis. The zenith angle estimator correctly identifies much of the
muon background as downgoing. Particle showers are preferentially interpreted as
either horizontally or diagonally passing tracks which allows for some
signal-background separation. 

{\bfseries Containment \& Vetoing:} 
A particularly problematic background are muon events which pass close to the
edges of the detector producing a light distribution which is similar to that
produced by cascades. To counter this background various containment conditions
are placed on the position of the reconstructed vertex, the first-hit DOM and
the DOM with the highest collected charge.

Events are excluded if the first hit DOM or the DOM with the highest collected
charge is located in the outermost vertical layer (see
Fig.~\ref{Figure:IC40detector}). Events are also vetoed if the depth of the
first hit DOM $z_{1^{\mathrm{st}}}$ falls in the top or bottom
$\unit{50}{\metre}$ of the detector.

The reconstructed vertex $\vec{x}_\textrm{CREDO}$ is required to lie inside the
instrumented volume and not in the top or bottom $\unit{50}{\metre}$ of the
detector. The analyses presented in this paper found different containment
conditions on the $xy$-position of the vertex to be optimal, when combined with
their particular selection cuts, for suppressing background while maintaining
signal efficiency. The two alternative containment conditions are illustrated by
the solid and dashed lines in Fig.~\ref{Figure:IC40detector}.

{\bfseries Topological Characteristics:} 
The hit patterns of particle showers in IceCube are approximately spherical
while the muon-track hit patterns are more elongated. A number of different
quantities can be calculated to characterize the different topology of cascade
and muon events and used to preferentially select cascades.

For example, a quantity analogous to the tensor of inertia of a rigid body, is
calculated for each event. The collected charge on each DOM takes the role of
the rigid body's mass distribution. The ratio of the smallest eigenvalue to the
sum of all of the eigenvalues, $\lambda = \lambda_\mathrm{min}/\sum\lambda_{i}$,
tends to $\frac{1}{3}$ for spherical events while muon tracks typically have
smaller eigenvalue ratios\cite{Ahrens:2003fg}. 

Another way to select spherical events is to construct a spherical volume
surrounding the reconstructed event vertex $\vec{x}_\textrm{CREDO}$ and consider
the proportion of hit DOMs, versus the total number of DOMS, in this sphere. The
radius of the sphere considered is chosen to scale with the average distance
between reconstructed vertex $\vec{x}_\textrm{CREDO}$ and position of the hit
DOMs -- a robust estimate for the overall size of the hit pattern. The
fill-ratio $f$ denotes the fraction of DOMs, falling within the sphere, on
which light is recorded\cite{Abbasi:2011ui}.  Hence, fill-ratios close to
$100\%$ are obtained for spherical hit patterns, while muon events yield lower
values. The fill-ratio is especially efficient to suppress coincident muon
events -- two or more muons from different air showers that cross the IceCube
detector within microseconds of each other. A second quantity, the difference
$\Delta f$ of two fill-ratios with different radii, is also used. This quantity
provides further separation power due to the fill-ratio's dependency on the
chosen radius being slightly different for the differently shaped hit patterns
of cascades and tracks.

{\bfseries Time Evolution and Charge Distribution:}
Below $\unit{}{\peta\electronvolt}$ energies the Cherenkov light of particle
showers originates within a few meters of the interaction vertex and then
propagates through the detector with the speed of light in ice $c_\textrm{ice}$.
In contrast, muon tracks traverse the detector with velocities close to the
speed of light in vacuum $c$ and emit Cherenkov photons continuously along their
track. Several approaches exploit this difference to separate cascade and track
events.

A simple approach, that can be applied before the event vertex has been
reconstructed, is the \texttt{linefit} algorithm. The hit pattern is fitted
with a straight line propagating with velocity
$v_\textrm{lf}$\cite{Ahrens:2003fg}. Relativistic muons often yield
\texttt{linefit} velocities close to $c$ whereas lower velocities are obtained
if the fit is applied to cascades.

If the interaction time and vertex are well reconstructed then causality can be
used to provide a strong constraint on whether an event is a neutrino-induced
particle shower event. For each DOM, light is expected to arrive at the earliest
after the time necessary to cover the distance between the DOM and the vertex,
at velocity $c_\textrm{ice}$. While delayed photon arrivals are common due to
light scattering in the ice, much earlier arrival times indicate a problem with
the cascade hypothesis. The difference between the expected
($t_\textrm{expected}$) and observed ($t_\textrm{hit}$) arrival time is
calculated for all DOMS and $\Delta t_\textrm{min}$ is defined as the smallest
such delay time $\Delta t_\textrm{min} =
\min(t_\textrm{hit}-t_\textrm{expected})$. Events with large negative values of
$\Delta t_\textrm{min}$ are removed.

Another approach is to sort the DOMs by the time they recorded light and then
consider the unit vector from one hit DOM to a subsequent hit DOM as an
individual dipole moment. The global dipole moment $m$ is obtained by averaging
over all individual dipole moments. Larger moments are expected from tracks and
smaller moments from cascades\cite{Ahrens:2003fg}.

Yet another way to emphasize the different hit pattern evolution of tracks and
cascades is to divide the event into two parts based on the times of the hit
DOMs. This splits tracks into two disjunct segments and cascades into two
mostly concentric shells. The \texttt{cscd\_llh} algorithm is used to obtain,
for each half, the vertices $\vec{x}_{1}$ and $\vec{x}_{2}$. Large radial and
vertical distances $\Delta r_{12}$ and $\Delta z_{12}$ between the vertices as
well as large differences in the reconstructed time $\Delta t_{12}$ are then
indicative of tracks. 

Another feature of cascade events, which is also related to the fact that the
particle showers are only a few meters in length, is that most of the light is
recorded close to the vertex and hence early in the event. In contrast, since
muons emit light continuously along their track it is more likely to see later
contributions to the total charge. The variable $\Delta t_{50\%-90\%}$ denotes
the fraction of the event length in which the total collected charge rises from
$50\%$ to $90\%$. Greater $\Delta t_{50\%-90\%}$ values indicate a longer time
interval for the second half of the event's total charge to be collected and are
more likely to occur for muon events. 

Some discrimination power comes from DOMs where just enough light arrives
to trigger the readout and hence only a single pulse is reconstructed from the
waveform. A combination of light yield, scattering in the ice and the geometric
shape of the hit pattern results in a slightly higher number of these DOMs for
muons. The ratio $n_{1} /n_{\mathrm{hit}}$ of DOMS with only one reconstructed
pulse over the total number of hit DOMs tends to smaller values for cascades.

Finally, a useful variable is the ratio of the charge collected in the DOM with
the highest charge, compared to the total recorded charge
$q_\textrm{max}/q_\textrm{tot}$. Low energetic muons passing very close to one
DOM can yield a high charge concentration in this DOM compared to others. These
events can resemble cascade-like hit patterns and are prone to be overestimated
in energy. Hence, requiring a low ratio $q_\textrm{max}/q_\textrm{tot}$ is
useful to reject this class of background events.


\subsection{Analysis overview}
Three event selections, named \EikeLE{}, \EikeHE{} and \Stephanie{}, were
developed to search for neutrinos in the IceCube-$40$ dataset. The multiple
event selections allowed sensitivity to both high-energy astrophysical and
low-energy atmospheric neutrinos as well as providing a collaboration internal
cross-check.

The different event selections rely on similar selection methods that are
differently combined for the individual goals of the analyses. The event
selections share the first three filter levels but differ at later filter
steps, since analysis \EikeLE{} aimed at measuring atmospheric neutrinos whereas
analyses \EikeHE{} and \Stephanie{} were optimized towards an astrophysical
neutrino flux. In order to avoid experimenter's bias a blind analysis was
performed. Each event selection was developed and tested on simulation and a
$10\%$ subset of the experimental data, sampled uniformly over the year. 
A simple cut-and-count experiment was done on the remaining $90\%$.

Analyses \EikeLE{} and \EikeHE{} were developed in parallel to analysis
\Stephanie{}, but only unblinded afterwards. The crucial difference is that they
utilized a significantly improved background simulation described in section
\ref{Section:Simulation}, resulting in the choice of tighter cuts and a purer
neutrino sample. Consequently here the focus is put on samples \EikeLE{} and
\EikeHE{}. An overview of the cuts performed to obtain the three samples is
given in Table~\ref{Table:CutComparison}. A comparison of the energy thresholds
and event rates is given in Table~\ref{Table:EikeUnblindingResults}.
 
Sizable systematic uncertainties must be considered in interpreting the result.
Therefore for samples \EikeLE{} and \EikeHE{} a Bayesian approach was chosen
(described in appendix \ref{Section:BayesianMethod}) and the result is reported
in form of the posterior probability for the number of non-background events.


\subsection{Filter Level 2 \& 3} \label{SubSection:OnlineFilterAndStraightCuts}
\begin{figure}[tb]
\subfigure[Line-fit velocity, $v_{lf} < 0.13$.]{
\includegraphics[width=0.49\textwidth]{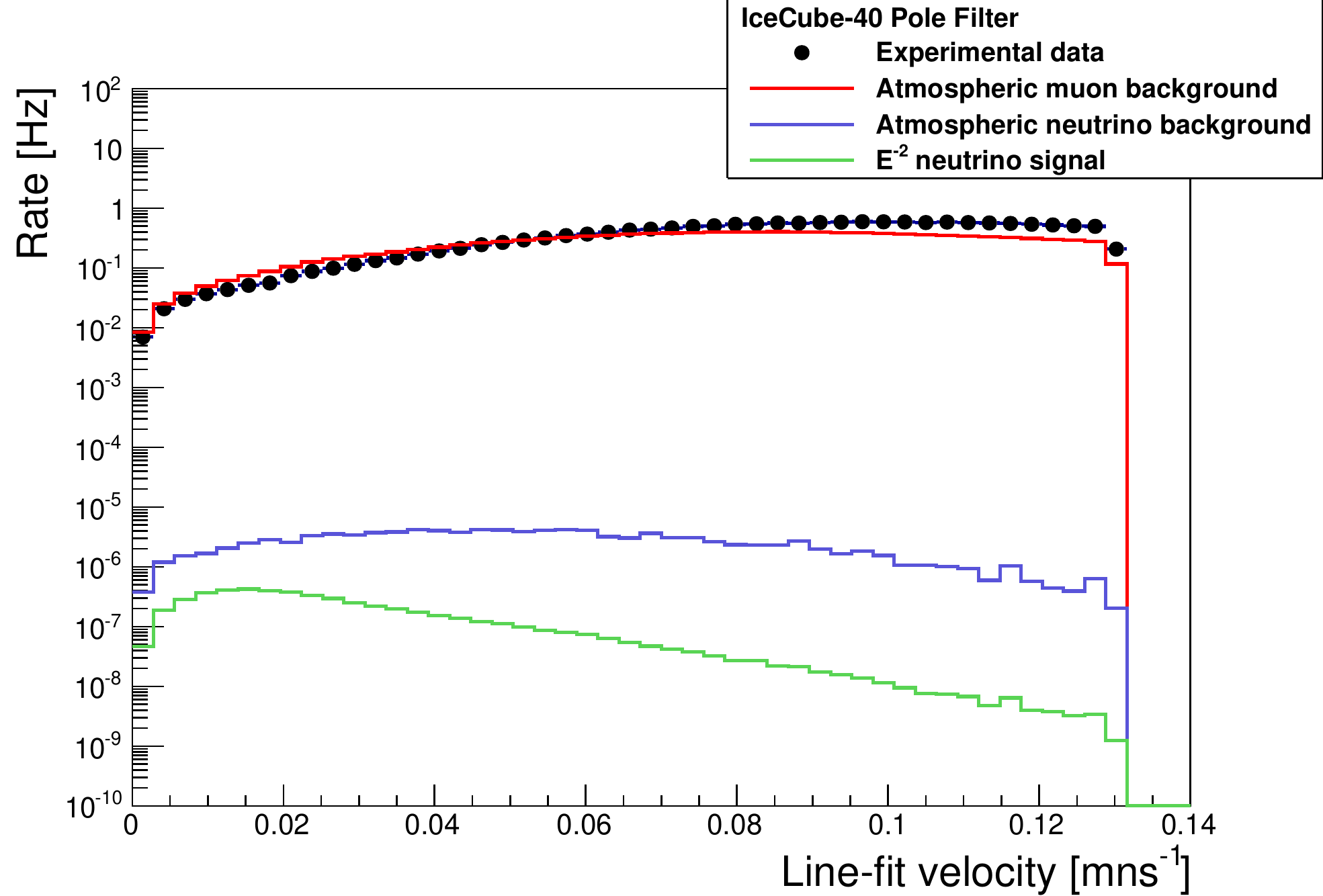}
\label{Subfigure:PoleFilter_1d_LFVel}}
\subfigure[Eigenvalue ratio, $\lambda > 0.12$.]{
\includegraphics[width=0.49\textwidth]{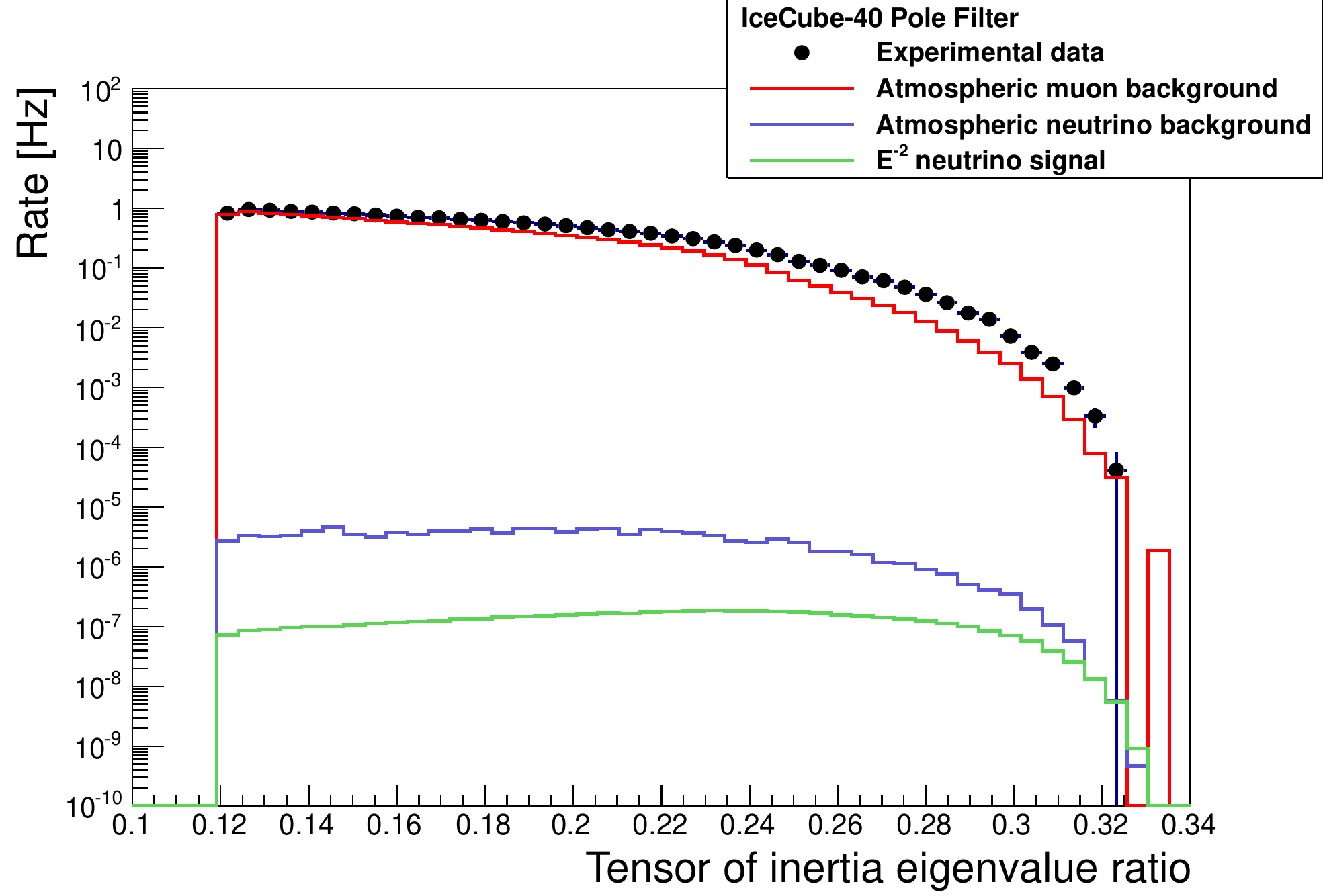}
\label{Subfigure:PoleFilter_1d_evalratio}}
\caption{The IceCube-$40$ online filter, the data is shown by the black filled
circles, simulated atmospheric muon background by the red line, simulated
atmospheric electron neutrinos by the blue line, and simulated electron neutrino
signal with spectrum $E^{2}dN/dE = \unit{3.6\times10^{-8}}{\flux}$ by the green
line.}
\end{figure}
\begin{figure}[tb]
\subfigure[The IceCube-$40$ Level $3$ zenith angle and energy reconstructions.
The two panels show the data and $E^{-2}$ spectrum electron neutrino signal. The
Level $3$ cuts are represented by the black lines at zenith
$\Theta_{\mathrm{track}} = \unit{80}{\degree}$ and energy $E_{\mathrm{ACER}} =
\unit{10}{\tera\electronvolt}$. Events in the hatched upper left quadrant of
each of the panels were removed.]{
\includegraphics[width=0.49\textwidth]{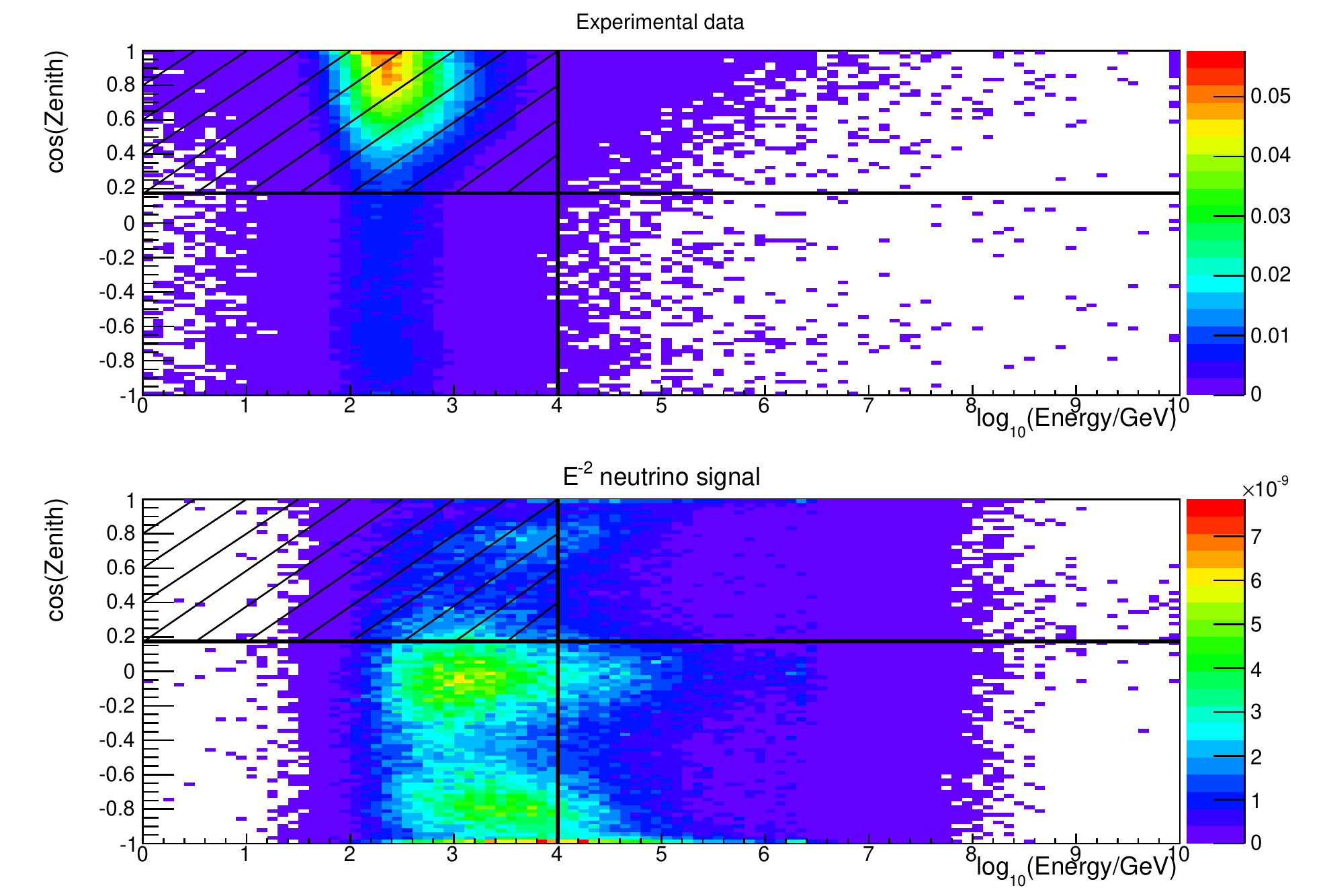}
\label{Figure:Level3_2d_Zenith_Energy}}
\subfigure[Reduced log-likelihood, $\mathrm{rlogL_{cscd}} < 10$]{
\includegraphics[width=0.49\textwidth]{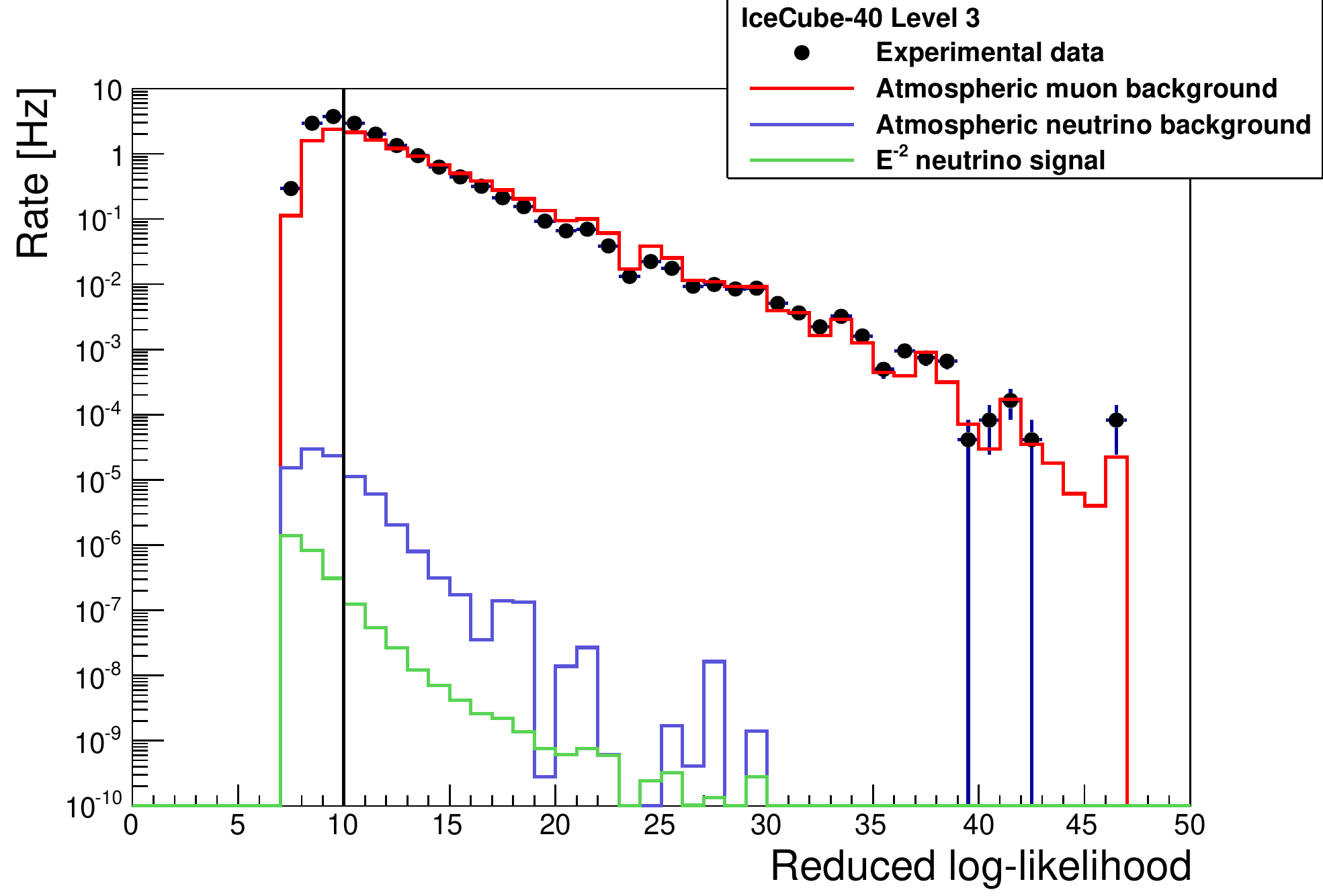}
\label{Figure:Level3_1d_reducedLlh}}
\caption{The IceCube-$40$ Level $3$ filter variables, the data is shown by the
black filled circles, simulated atmospheric muon background by the red line,
simulated atmospheric electron neutrinos by the blue line, and simulated
electron neutrino signal with spectrum $E^{2}dN/dE =
\unit{3.6\times10^{-8}}{\flux}$ by the green line.}
\end{figure}

Triggered events are filtered online at the South Pole to reduce the data volume
to the bandwidth available for data transfer from the Pole via satellite. The
online filter, or Level $2$ is a filter on easy-to-compute variables that
retains the majority of the neutrino signal while removing a large fraction of
the background. It requires the \texttt{linefit} velocity to be $v_\textrm{lf} <
\unit{0.13}{\metre/\nano\second}$ and the tensor-of-inertia eigenvalue ratio to
be $\lambda > 0.12$. The distributions of these variables are shown in
Figs.~\ref{Subfigure:PoleFilter_1d_LFVel} and
\ref{Subfigure:PoleFilter_1d_evalratio}. At this early stage of the event
selection, where no event quality cuts are done yet, the distributions show
some disagreement between data and simulation. Further steps in the event
selection will reduce this disagreement, mostly by removing events that are too
low in energy or that are not contained in the instrumented volume. The pole
filter reduced the data rate by two orders of magnitude to approximately
$\unit{16}{\hertz}$. 

Due to the energy threshold of about $\unit{100}{\giga\electronvolt}$ and the
steeply falling energy spectrum, the energy distribution of background events,
passing Level 2 is strongly peaked at lower energies. As the majority of the
events are still atmospheric muons most of the events are reconstructed as
downgoing tracks (see Fig.~\ref{Figure:Level3_2d_Zenith_Energy}). Consequently,
the Level $3$ cuts concentrate on reducing the background at low energies. 

All events with a reconstructed energy $E_\textrm{ACER} >
\unit{10}{\tera\electronvolt}$ are kept, whereas lower energetic events are
subject to two more cuts. First, when reconstructed under a muon track
hypothesis, events reconstructed as downgoing with zenith angles
$\Theta_\textrm{track} < \unit{80}{\degree}$ are rejected. Secondly, those
events are removed, where a \texttt{cscd\_llh} likelihood value of
$\textrm{rlogL}_\textrm{cscd} > 10$ indicates a poorly fitted event. After
application of the Level $3$ filter the data rate is reduced by another order
of magnitude to approximately $\unit{1.8}{\hertz}$, while contained
astrophysical (atmospheric) neutrinos were kept with an efficiency of $76.5\%$
($56.6\%$).


\subsection{Sample \EikeLE{}}
After Level $3$, the analyses diverge. Event selection \EikeLE{} aims at the
observation of atmospheric neutrinos with energies of a few
$\unit{}{\tera\electronvolt}$. Previous cuts aimed to provide an optimal
efficiency for cascades regardless of their energy and position inside the
detector. As a consequence at Level $3$ a significant number of events, that
are very low in energy or happening at the border of the instrumented volume,
remain in the sample. For these events no reliable separation between signal
and background could be found.

Therefore, a set of geometric conditions has been defined to remove such events:
it is required that the first hit DOM is neither on the outer layer of
IceCube-$40$ (red squares in Fig.~\ref{Figure:IC40detector}) nor in the top or
bottom $\unit{50}{m}$ of the detector. Also DOMs on at least five different
strings must have recorded light. This cut reduces the data rate to
$\unit{79}{\milli\hertz}$, while contained astrophysical (atmospheric)
neutrinos are kept with an efficiency of $56.1\%$ ($37.2\%$).

After these cuts, at the so-called Level $5$, additional reconstructions but no
cuts were performed. In particular the remaining sample was small enough to
perform the \texttt{CREDO} likelihood reconstruction described in section
\ref{Section:ReconstructionOfParticleShowers} with $8$ iterations. Muon
background still dominates over atmospheric neutrino-induced showers by three
orders of magnitude at selection Level $5$. Therefore, the vertex and energy
estimates are used to apply another series of event selection conditions.

\begin{figure*}[tb]
\includegraphics[width=0.75\textwidth]{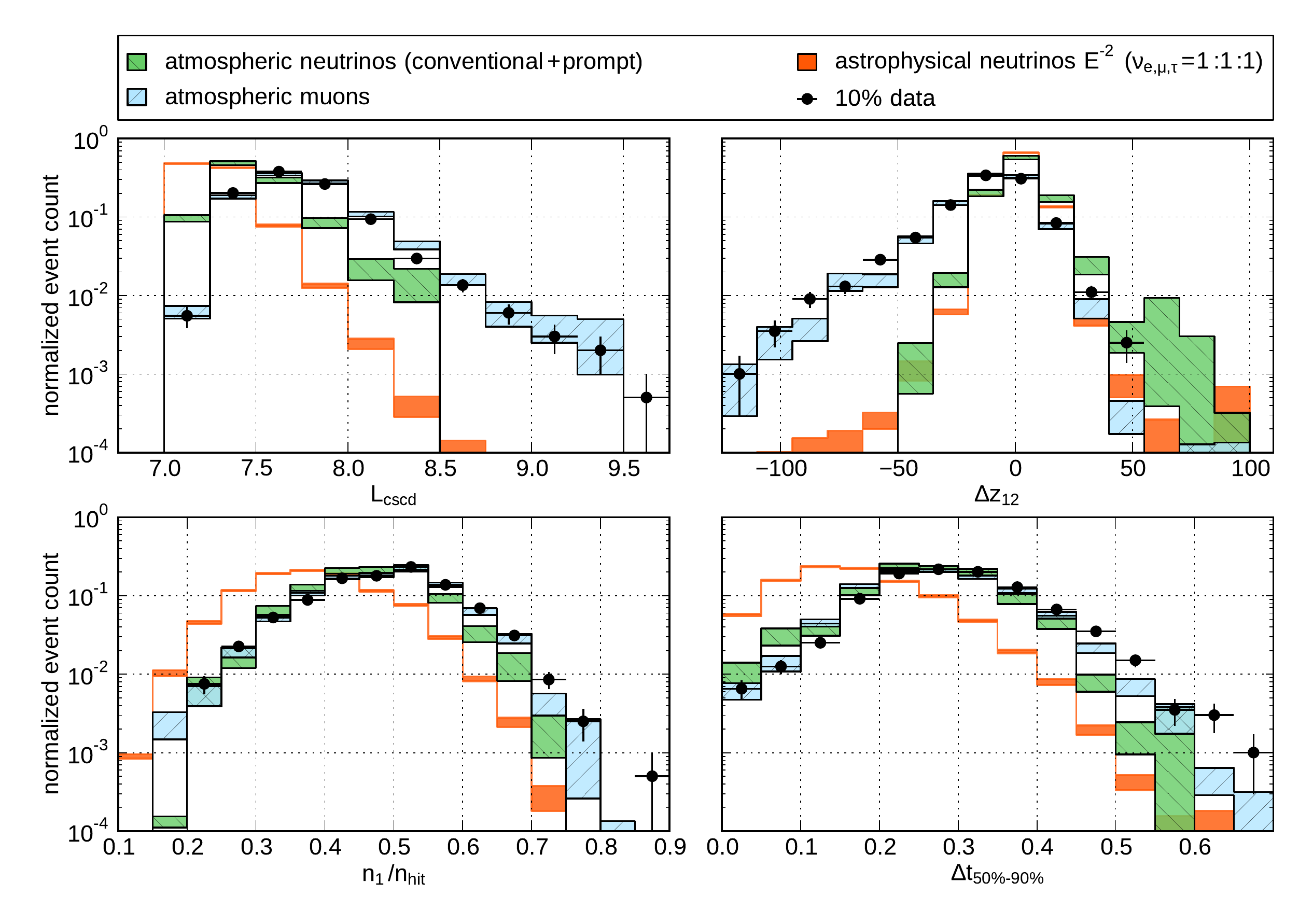}
\caption{At Level $6$ eleven input variables are combined to train a boosted
decision tree. Individually the variables provide only limited separation power,
especially for distinguishing atmospheric neutrinos and muons. The four most
important variables according to the classifier, are shown here at Level $6$.
Definitions of the variables can be found in section \ref{Section:Variables}. 
The shaded regions in the plot denote the statistical uncertainties.}
\label{Figure:TMVAEike_inputVariables}
\end{figure*}

At Level $6$, the energy threshold is set to $E_\textrm{CREDO}^{(8)} >
\unit{1.8}{\tera\electronvolt}$. In order to remove events where low energetic
muons pass very close to a DOM and where their energy is likely to be
overestimated, a cut on $q_\textrm{max}/q_\textrm{tot} < 0.3$ is applied.
Furthermore, the containment requirement is enforced: the reconstructed vertex
must lie inside the blue solid boundary shown in Fig.~\ref{Figure:IC40detector},
its $z$-coordinate must be inside the instrumented depth interval and also the
DOM with the highest recorded charge in the event may not be on the outer layer.
A constraint on the minimum delay time $\Delta t_\textrm{min} >
\unit{-75}{\nano\second}$ is used to remove events where causality rules out the
cascade hypothesis. Together with requiring a fill-ratio $f > 0.6$, this set of
cuts reduces the data rate by two orders of magnitude down to
$\unit{660}{\micro\hertz}$, while retaining efficiencies for contained
astrophysical and atmospheric neutrinos of $66.4\%$ and $20.1\%$, respectively.
Overall agreement of event properties between simulation and experimental data
is good (see Fig.~\ref{Figure:TMVAEike_inputVariables}). This is a necessary
precondition for the final event selection step, which uses a multivariate
algorithm to distinguish signal and background events.

The final step of the event selection is to remove the remaining background and
obtain a sample dominated by neutrinos. For this purpose event properties that
still provide separation power are used as input to a multivariate algorithm
to obtain a single quality parameter for each event. The TMVA
package\cite{Hocker:2007zz} is used to train a boosted decision tree (BDT). In
total $11$ variables are combined into one final event quality variable
$\BDTII$. These are the \texttt{cscd\_llh} algorithm's quality parameter
$\textrm{rlogL}_\textrm{cscd}$, the vertical and radial distances $\Delta
z_{12}, \Delta r_{12}$ of the time-split reconstructions, the fraction of DOMs
with only one reconstructed pulse $n_{1} /n_{\mathrm{hit}}$ and the fraction of
the event duration needed to accumulate the second half of the total charge
$\Delta t_{50\%-90\%}$. The two zenith angle estimators
$\Theta_\textrm{CREDO}^{(8)}$ and $\Theta_\textrm{track}$ enter the BDT as well
as the dipole moment $m$ and the fill-ratio difference for two radii $\Delta f$.
Finally, the eigenvalue ratio $\lambda$ and the minimum delay time $\Delta
t_\textrm{min}$ are used, too. They have already been used at earlier cut levels
but still provide some discrimination power. The four variables ranked most
important by the classifier are shown in
Fig.~\ref{Figure:TMVAEike_inputVariables}.

\begin{figure}[tb]
\includegraphics[width=0.49\textwidth]{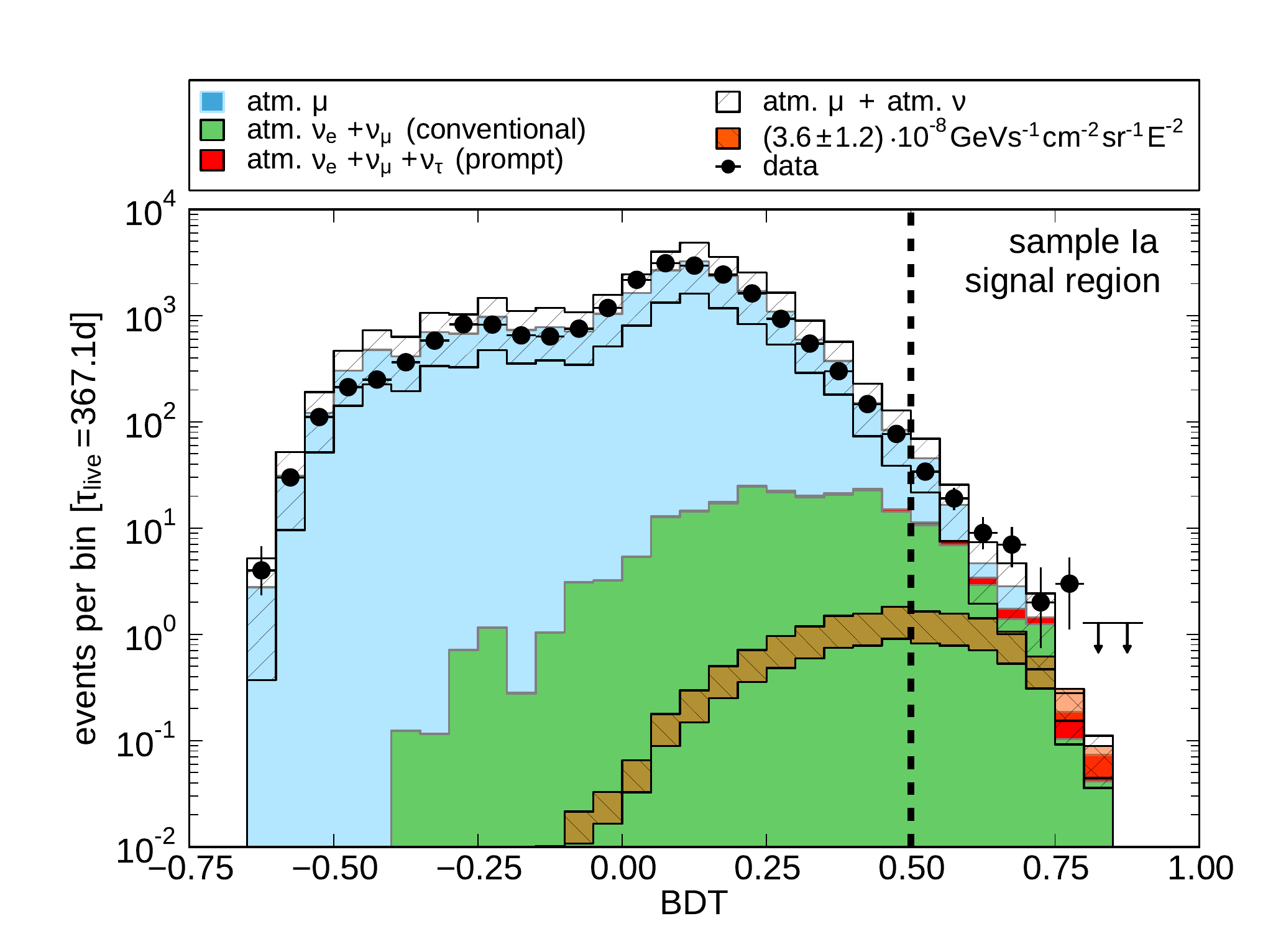}
\caption{The BDT output variable at Level $6$. A cut at $\BDTII > 0.5$ defines
the sample \EikeLE{}. The atmospheric muon and neutrino components are stacked
on top of each other and describe together the recorded data well. The
experimental data shown comprises $100\%$ of the sample.  For bins in which no
event was observerd a $68\%$ C.L. upper limit is shown.  The white hatched area
shows the distribution of atmospheric muons and neutrinos, including systematic
and statistical uncertainties. The orange hatched area denotes the prediction
of astrophysical neutrinos according to the flux estimate from\cite{hese}.}
\label{Figure:EikeBDT}
\end{figure}

The distribution of the BDT output variable and its ability to separate signal
from background is shown in Fig.~\ref{Figure:EikeBDT}. At low values, the
distribution is dominated by the atmospheric muon background and the
experimental data is adequately described by the \texttt{CORSIKA} simulation.
The signal distribution centers at higher BDT scores, but even there a
contribution from atmospheric muons is present. Closer inspection reveals that
these events are muons with prominent bremsstrahlung cascades and little to no
hint of the muon track. The events resemble neutrino-induced particle showers
rather well. The geometry of IceCube-$40$ with one dimension being much shorter
than the others is obviously vulnerable to this class of background events. In
the energy range below $\unit{100}{\tera\electronvolt}$, with the given detector
and the aforementioned background rejection methods, including their combination
within a multivariate classifier, these events turn out to be irreducible
background. 

This limits the possible background suppression for sample \EikeLE{}.
Atmospheric neutrino signal and muon background are similarly distributed at
$\BDTII$ scores $> 0.5$, so cutting at a higher value removes the atmospheric
neutrino signal as much as the atmospheric muon background. As a consequence a
$\BDTII$ value of $0.5$ is the optimal separation point between background and
signal that maximizes the detection potential\cite{Hill:2002nv} for an
atmospheric neutrino flux. However, a rather large number of $71$ events is
expected in sample \EikeLE{} from which $41$ are expected to be atmospheric
muons and $30$ conventional and prompt neutrinos. Both numbers are affected by
rather large systematic uncertainties, which are discussed in section
\ref{Section:SystematicUncertainties}.

\subsection{Sample \EikeHE{}} \label{SubSection:SampleEikeHE}
\begin{figure}[tb]
\includegraphics[width=0.49\textwidth]{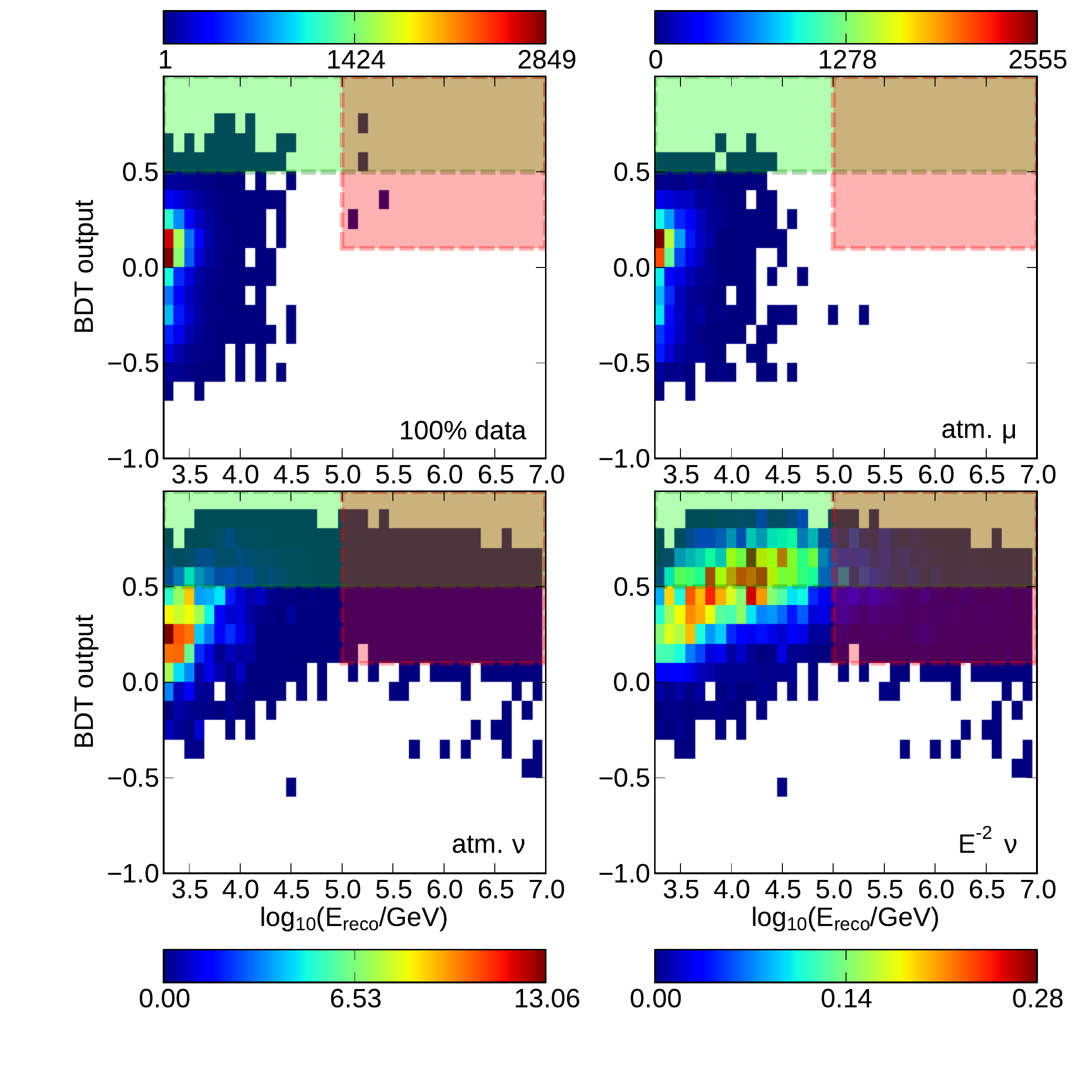}
\caption{The parameter space spanned by the BDT output variable and the
reconstructed energy. The green shaded area ($\BDTII > 0.5$) and the red shaded
area ($\BDTII > 0.1) \& (\log_{10}(E_{\textrm{reco}}/\unit{}{\giga\electronvolt})
> 5$) denote the cuts of samples \EikeLE{} and \EikeHE{}, respectively. The color
scale denotes linearly the expected/seen number of events per bin in $100\%$ of
the livetime.}
\label{Figure:EikeParspace}
\end{figure}

The remaining background events can be removed by increasing the energy
threshold of the analysis. Because of their steeply falling spectrum
conventional atmospheric neutrinos are severely reduced by an energy cut.
However, the harder spectra of prompt neutrinos and the assumed diffuse
astrophysical neutrino flux are not so much affected. This fact is used to
define the data sample \EikeHE{} with an energy threshold of
$E_\textrm{CREDO}^{(8)} > \unit{100}{\tera\electronvolt}$. A comparatively loose
requirement on the BDT score of $\BDTII > 0.1$ is then sufficient to remove all
simulated background events (see Fig.~\ref{Figure:EikeParspace}). In contrast to
sample \EikeLE{}, which was designed to find atmospheric neutrinos, sample
\EikeHE{} has a significantly better performance for detecting a prompt or
astrophysical neutrino flux.

For this sample, the background contribution from atmospheric muons cannot be
determined from simulation anymore as all events have been removed. 
In order to estimate the muon rate, the reconstructed energy distribution is
extrapolated from the low-energy region, where simulated events are still
available, to higher energies.

Up to $\unit{}{\peta\electronvolt}$ energies, the energy spectrum of the
brightest bremsstrahlung cascades along simulated muon tracks can be described
by a power-law $dN/dE \propto E^{-3.7}$. For muon events, which pass the event
selection, the corresponding energy spectrum is not necessarily the same,
because of the energy-dependent performance of the cuts. E.g. muon events with
bright bremsstrahlungs cascades are suppressed with increasing energy, since
the muon becomes more likely to be detected in the veto region. On the other
hand, the selection efficiency for particle showers rises with increasing
deposited energy. Hence, the model used to extrapolate the muon background into
the signal region is the product of two functions: the selection efficiency for
particle showers as a function of deposited energy (derived from simulation)
and a power-law with the index and normalization as free parameters.

\begin{figure}[tb]
\subfigure[Deposited Energy in sample \EikeLE{}]{
\includegraphics[width=0.49\textwidth]{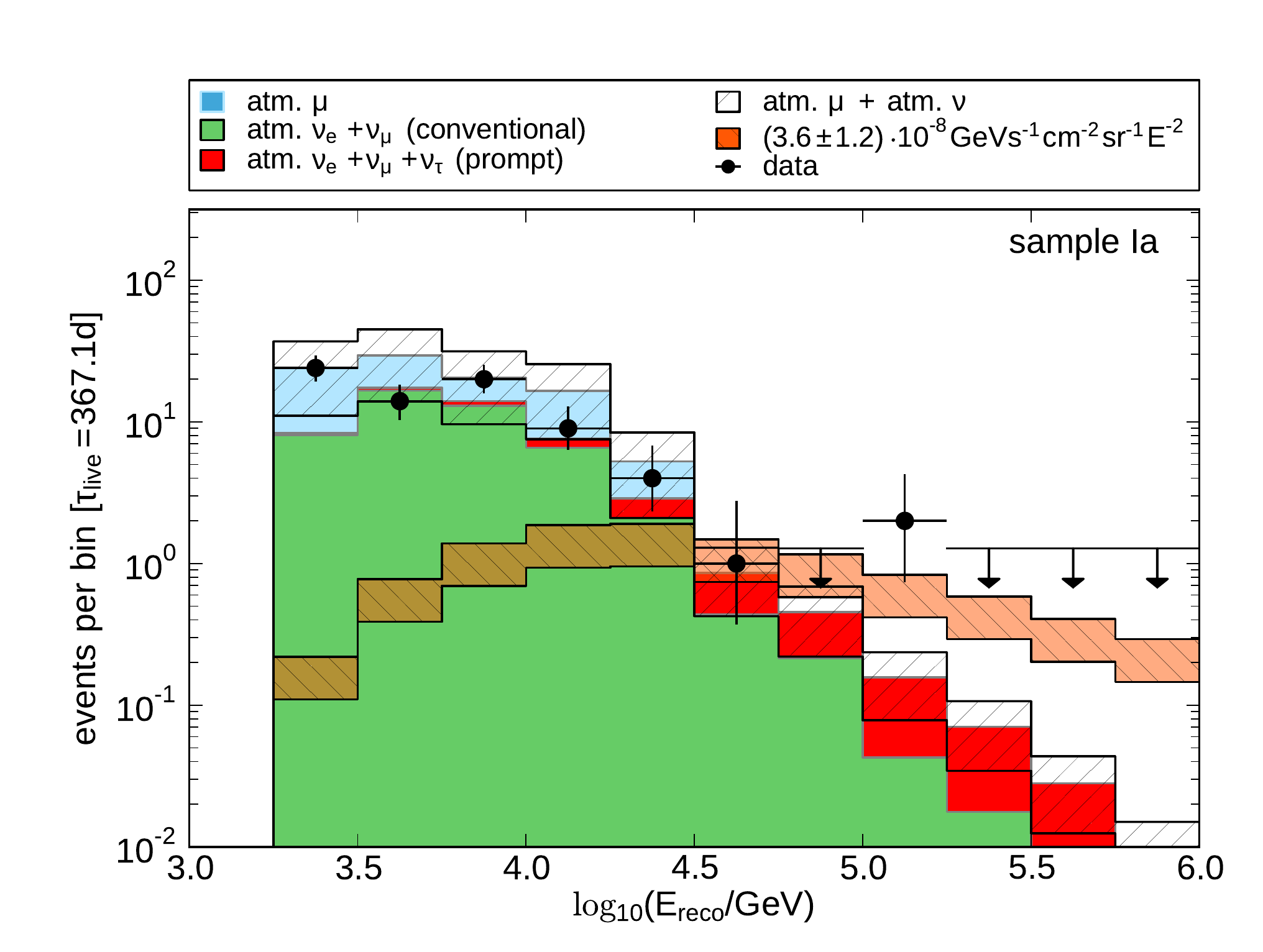}
\label{Figure:EikelowE}}
\subfigure[Deposited Energy in sample \EikeHE{}. The plot extends into the
background region below $\unit{10^5}{\giga\electronvolt}$. From there the
contribution of atmospheric muons is extrapolated to higher energies (black
dashed line).]{
\includegraphics[width=0.49\textwidth]{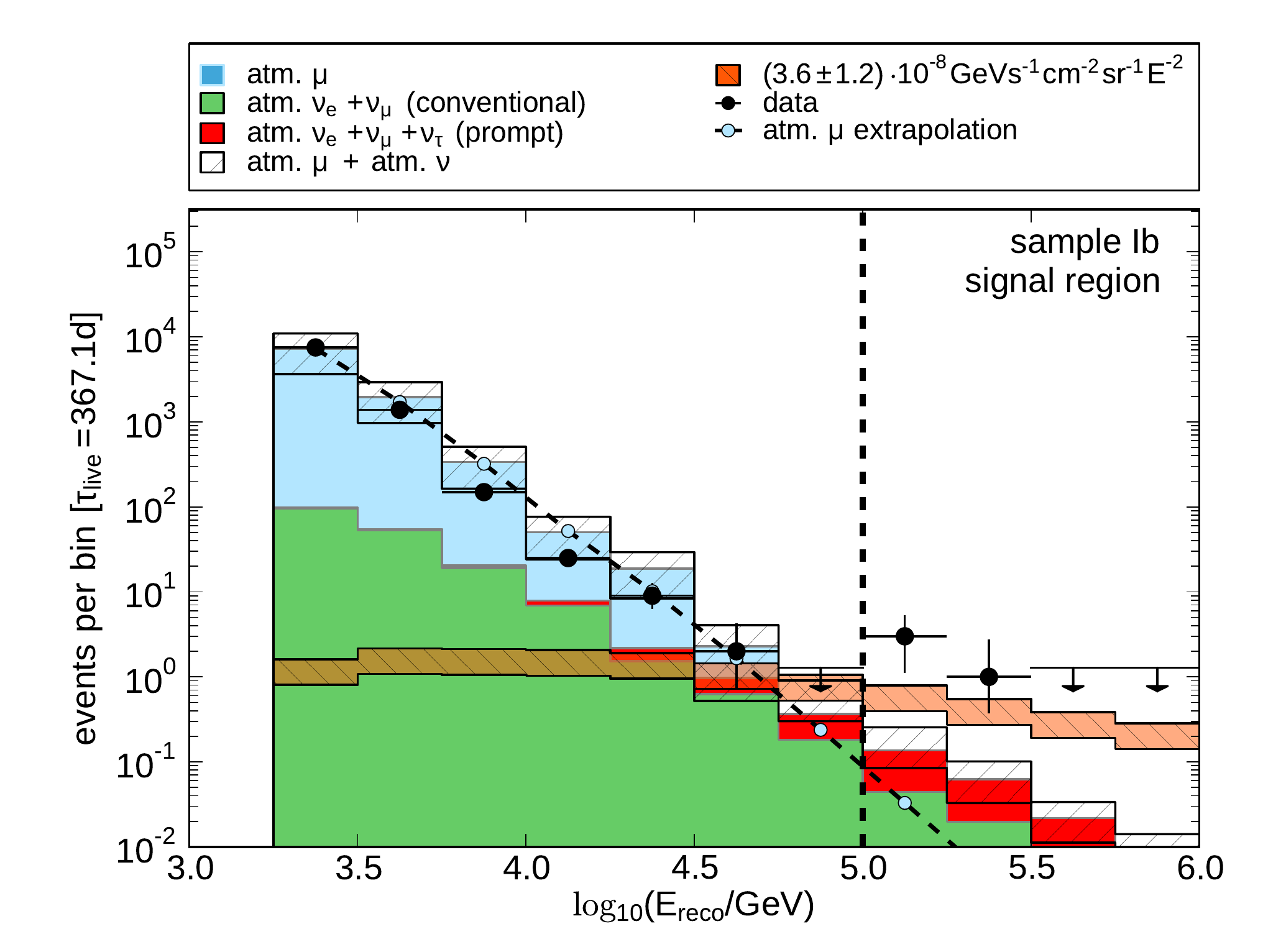}
\label{Figure:EikehighE}}
\caption{ The reconstructed energy distributions for the samples \EikeLE{} (a) and
\EikeHE{} (b). 
The atmospheric muon and neutrino components
are stacked on top of each other. The experimental data shown comprises $100\%$
of the sample. For bins in which no event was observerd a $68\%$ C.L. upper
limit is shown. The white hatched area shows the distribution of atmospheric
muons and neutrinos, including systematic and statistical uncertainties. The
orange hatched area denotes the prediction of astrophysical neutrinos according
to the flux estimate from\cite{hese}.}
\end{figure}

This model provides a reasonable fit to the energy distribution of background
muons below the energy threshold (see Fig.~\ref{Figure:EikehighE}, the same
$\BDTII$ score $> 0.1$ is required for these events as for the events in the
high-energy sample). The fit is consistent with the expectation that the
remaining background events are dominated by high bremsstrahlung cascades and
that the energy estimator describes their energy spectrum. This confirms the
physical motivation of the extrapolation in the signal region. From the
extrapolation one can amount the muon contribution to $0.04^{+0.06}_{-0.02}$
events, where the errors are derived from varying the parameters within the
uncertainties obtained from the fit. An additional $0.21$ events are expected
from conventional and prompt atmospheric neutrinos.

\begin{figure}[tb]
\includegraphics[width=0.49\textwidth]{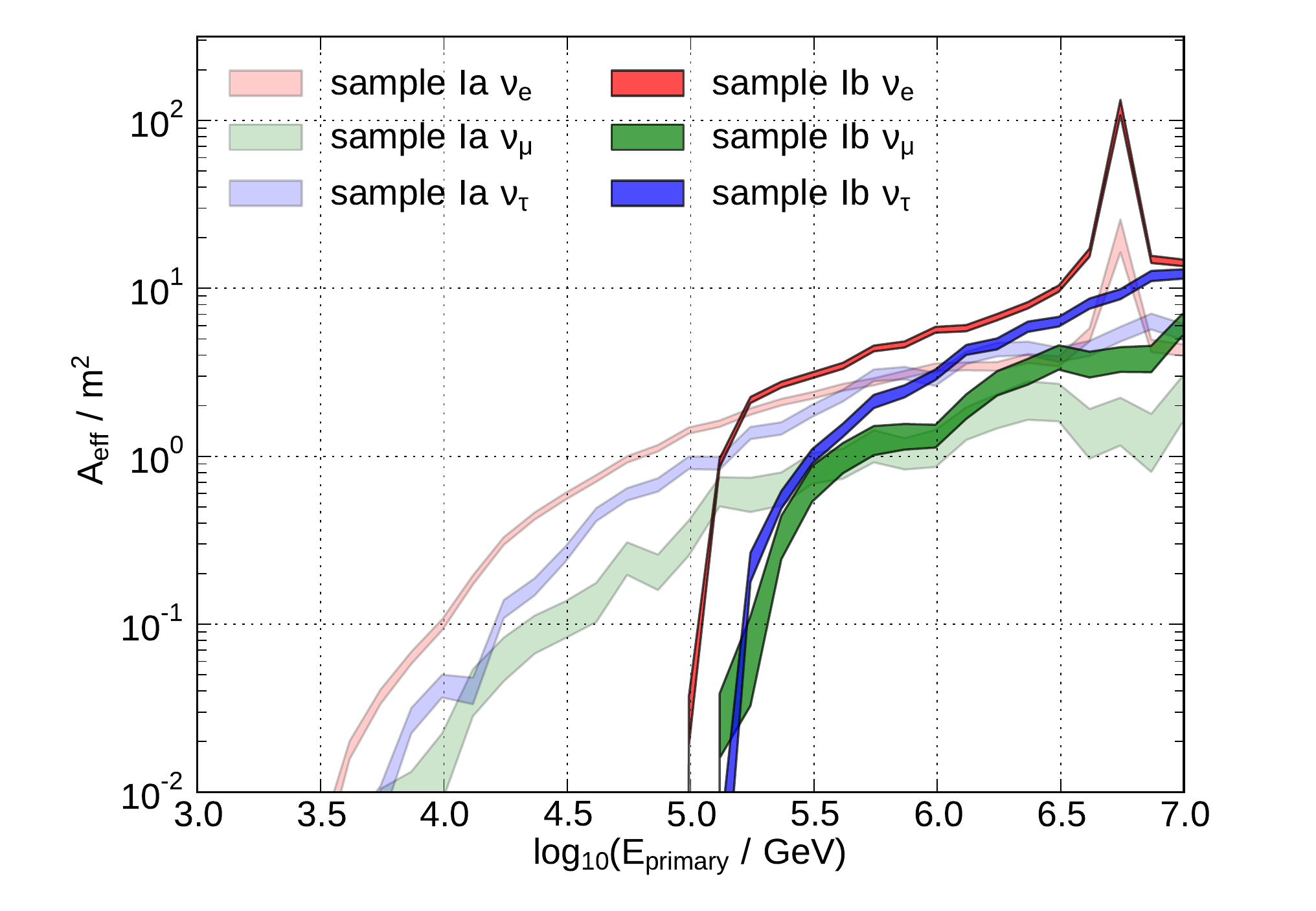}
\caption{Effective areas for different neutrino flavors in samples \EikeLE{} and
\EikeHE{}.}
\label{Figure:EikeEffectiveArea}
\end{figure}

This event selection reduced the muon background by seven orders of magnitude
from the rate after the online filter. This high suppression comes at the price
of a low total efficiency of $5\%$ ($0.5\%$) for contained astrophysical
(atmospheric) neutrinos. The energy-dependent selection efficiencies are
presented in the form of effective areas for both samples in
Fig.~\ref{Figure:EikeEffectiveArea}. Table~\ref{Table:EikeEfficiencies}
summarizes the performance of the event selection at different cut levels. 

\begingroup
\squeezetable
\begin{table}[t]
\begin{ruledtabular}
\begin{tabular}{@{}l|ccc@{}}
\multirow{2}{*}{Cut Level} & Experimental & \multicolumn{2}{c}{Contained neutrinos}                     \\
                           & data                  & $3.6\times 10^{-8} E^{-2}$  & HKKMS07 (no knee)    \\ \hline
                           &\multicolumn{3}{c}{Rates in $\unit{}{\micro\hertz}$}                        \\ 
Level 2                    & $16.03\times10^6$     & $1.28 \pm 0.03$             & $153 \pm 4$          \\
Level 3                    & $1.74\times10^6$      & $0.97 \pm 0.01$             & $86 \pm 1$           \\
Level 4/5                  & $78980$               & $0.550 \pm 0.008$           & $32.1 \pm 0.7$       \\
Level 6                    & $660$                 & $0.365 \pm 0.006$           & $6.5 \pm 0.3$        \\
Sample \EikeLE{}           & $2.34$                & $0.156 \pm 0.004$           & $0.73 \pm 0.08$      \\
Sample \EikeHE{}           & $0.105$               & $0.068 \pm 0.001$           & $0.0021 \pm 0.0003$  \\ \hline
                           & \multicolumn{3}{c}{Efficiencies with respect to previous level}            \\ 
Level 3                    & $10.8\%$              & $76.5\%$                    & $56.6\%$             \\
Level 4/5                  & $4.5\%$               & $56.1\%$                    & $37.2\%$             \\
Level 6                    & $0.8\%$               & $66.4\%$                    & $20.1\%$             \\
Sample \EikeLE{}           & $0.4\%$               & $42.8\%$                    & $11.3\%$             \\
Sample \EikeHE{}           & $0.02\%$              & $18.6\%$                    & $0.04\%$             \\
\end{tabular}
\caption{Event rates and cut efficiencies at the different levels leading to
samples \EikeLE{} and \EikeHE{}. While the data column refers to all recorded
events that pass the cuts, the neutrino rates refer to contained events, i.e.
neutrinos which have their interaction vertex inside the area circumscribed by
the blue solid line in Fig.~\ref{Figure:IC40detector}. For charged current
$\nu_\mu$ interactions the muon must have its highest energy loss inside the
area.}
\label{Table:EikeEfficiencies}
\end{ruledtabular}
\end{table}
\endgroup

\begin{center}
\begin{table*}[t]
{\small
\hfill{}
\begin{tabular}{|c|p{4cm}|p{4cm}|p{6.5cm}|}
\hline
Optimized for & \hfil Atmospheric neutrinos\hfil & \multicolumn{2}{|c|}{Astrophysical neutrinos} \\ \hline 
              & \hfil Sample \EikeLE{} \hfil  & \hfil Sample \EikeHE{} \hfil & \hfil Sample \Stephanie{}  \hfil \\ \hline
Level $1$     & \multicolumn{3}{|c|}{light recorded on $8$ different DOMs within a $\unit{5}{\micro\second}$ time window} \\ \hline
Level $2$     & \multicolumn{3}{|c|}{ $v_\textrm{lf} < \unit{0.13}{\metre/\nano\second}$, $\lambda > 0.12$} \\\hline

Level $3$     & \multicolumn{3}{|c|}{ ($E_{\mathrm{ACER}} > \unit{10}{\tera\electronvolt}$) or ($E_{\mathrm{ACER}} < \unit{10}{\tera\electronvolt}$, 
               $\Theta_{\mathrm{track}} > 80^\circ$, $\textrm{rlogL}_\textrm{cscd} < 10$)}\\ \hline

Level $4$     & \multicolumn{2}{|p{8cm}|}{ $N_{\mathrm{strings}} > 5$, $\unit{-450}{\metre} < z_{1^{\mathrm{st}}} < \unit{+450}{\metre}$, \newline 
               DOM with first hit not on outer string} 
              & $E_\textrm{CREDO}^{(1)} > \unit{2.5}{\tera\electronvolt}$, $\Delta r_{12} < \unit{40}{\metre}$,\newline $f > 0.4$  \\ \hline

Level $5$/$6$ & \multicolumn{2}{|p{8cm}|}{DOM with largest charge not on outer string, \newline $\vec{x}_\texttt{CREDO}^{(8)}$ contained, 
               $\unit{-500}{\metre} < z_\texttt{CREDO}^{(8)} < \unit{+500}{\metre}$, \newline
               $E_\textrm{CREDO}^{(8)} > \unit{1.8}{\tera\electronvolt}$, $\Delta t_\textrm{min} > \unit{-75}{\nano\second}$, \newline
               $q_{\mathrm{max}}/q_{\mathrm{total}} < 0.3$, $f > 0.6$} 
              & DOM with largest charge not on outer string, 
               $\vec{x}_\texttt{CREDO}^{(4)}$ contained, \newline
               $\unit{-450}{\metre} < z_\texttt{CREDO}^{(4)} < \unit{+450}{\metre}$ \\ \hline

BDT input  
              & \multicolumn{2}{|p{8cm}|}{$\textrm{rlogL}_\textrm{cscd}, \Delta r_{12}, \Delta z_{12}, n_{1}/n_\mathrm{hit}, 
                                      \Delta t_{50\%-90\%}, \Delta t_\textrm{min},$ \newline
                                      $\cos(\Theta_\textrm{CREDO}^{(8)}),\cos(\Theta_\textrm{track}),  \Delta f, \lambda, m $ } 
              & $z_\texttt{CREDO}^{(4)}, \Theta_{\mathrm{track}},\textrm{rlogL}_\textrm{track}, v_{\mathrm{lf}},$\newline 
               $\lambda, f,\Delta t_{12},|\vec{x}_1|$ \\ \hline

Final Cuts    & $\BDTII > 0.5$ 
              & $\BDTII > 0.1$ \newline 
               $E_{\mathrm{CREDO}}^{(8)} > \unit{100}{\tera\electronvolt}$ 
              & $\BDTI > 0.2$, \newline $E_{\mathrm{CREDO}}^{(4)} > \unit{25}{\tera\electronvolt}$ \\ \hline
\end{tabular}}
\hfill{}
\caption{Comparison of the event selections. The symbols used here are described
in section \ref{Section:Variables}.}
\label{Table:CutComparison}
\end{table*}
\end{center}


\subsection{Sample \Stephanie{}} \label{SubSection:SampleStephanie}
Level $4$ for the event selection sample \Stephanie{} enforces a moderate energy
cut of $E_\textrm{CREDO}^{(1)} > \unit{2.5}{\tera\electronvolt}$ and selection
for cascade-like events using two of the topological variables described in
Section \ref{Section:Variables}. The difference between the split vertex reconstruction
is restricted to ($\Delta r_{12} < \unit{40}{\metre}$) and high fill-ratios
($f > 0.4$) are required. The \texttt{CREDO} reconstruction is performed with
four iterations on the remaining events to increase the accuracy of the vertex
and energy estimate. The next set of cuts focus on selecting contained events.
The $xy$-coordinates of $\vec{x}_\texttt{CREDO}^{(4)}$ must lie inside the area
circumscribed by the blue dashed line in Fig.~\ref{Figure:IC40detector} and the
$z$-coordinate must lie between $\unit{\pm 450}{\metre}$. The DOM with the
highest charge is not allowed to be on the veto layer of strings.

As in samples \EikeLE{} and \EikeHE{}, a boosted decision tree is trained to 
combine the variables, which still have discrimination power, into one event
quality parameter $\BDTI$. Eight variables are used: the zenith angle
$\Theta_\textrm{track}$ and fit quality parameter
$\textrm{rlogL}_\textrm{track}$ of the track reconstruction; the online filter
variables $v_\textrm{lf}$ and $\lambda$; the fill-ratio $f$ and the vertex time
difference $\Delta t_{12}$. Information on where in the detector the event
occurred entered the BDT in the form of the $z$-coordinate of
$\vec{x}_\texttt{CREDO}^{(4)}$ and the distance of $\vec{x}_1$ to the detector
center. A cut at $\BDTI > 0.2$ and
$E_\texttt{CREDO}^{(4)}>\unit{25}{\tera\electronvolt}$ is found to be optimal
to select astrophysical neutrinos.

The event selection criteria for sample \Stephanie{} were finalized before those
of sample \EikeLE{} and \EikeHE{}. The available statistics of simulated
background events were limited and did not describe the data particularly well
due to a lack of high-energy proton-induced air showers. Hence, unlike for
samples \EikeLE{} and \EikeHE{}, the full simulation discussed in section
\ref{Section:Simulation} did not enter the optimization of the event selection.
Instead, when training the BDT, experimental data was used for the background
sample. The final energy threshold was set to $\unit{25}{\tera\electronvolt}$,
which---given the limited background simulation available at that time---lead to
an optimal sensitivity for an astrophysical $E^{-2}$ flux. The final numbers
presented here for the background estimate are based on the full statistics.
Further information about the event selection can be found
in\cite{Hickford:2011}. 

\section{Systematic Uncertainties} \label{Section:SystematicUncertainties}
The three event samples (\EikeLE{}, \EikeHE{}, \Stephanie{}) share common
sources of systematic uncertainties. The largest uncertainties on the expected
event count arise from our limited knowledge of the optical ice properties at
the South Pole, the selection efficiency of the muon background and the
theoretical predictions for the atmospheric neutrino flux. Table
\ref{Table:EikeSystematics} summarizes the systematic uncertainties for sample
\EikeLE{} and \EikeHE{}. The systematic uncertainties associated with sample
\Stephanie{} are similar to those of sample \EikeHE{}. The individual systematic
errors are described in more detail below.

In calibration runs, where LEDs inside the DOMs illuminate the detector, data is
obtained from which the characteristics of the light absorption and scattering
at different points throughout the ice can be estimated. The estimated
characteristics form a so-called ice model which contains the depth, wavelength,
and temperature dependent optical properties throughout the detector and the
surrounding ice and bedrock. The AHA ice model\cite{Ackermann:2006} was used in all
simulations for this analysis. To carry out a study of the uncertainties arising
from the ice model, simulated datasets were produced using an alternative ice
model, called SPICE\cite{Abbasi:2011zz} which was developed after this analysis
was complete. The systematic uncertainty from the ice models was estimated to be
$\pm24\%$ for an $E^{-2}$ neutrino spectrum signal and $\pm11\%$ for the
atmospheric neutrino background for the three samples \EikeLE{}, \EikeHE{} and
\Stephanie{}. 

The estimate of the rate of atmospheric muons that pass the selection cuts is
also affected by a rather large uncertainty. The simulation falls short in
providing an absolute rate estimate, by systematically underestimating the
measured muon rate. At cut levels before Level $6$, where the data-to-simulation
agreement is not yet optimal, cuts affect data and simulation events slightly
differently. Hence, the rate discrepancy changes slightly between cut levels. At
each of these levels the rate estimate from atmospheric muon simulation can be
normalized to match the measured data rate by applying factors of $1.25$--$2$.
As described in section \ref{Section:EventSelectionAndAnalysisMethod}, the event
selection for sample \EikeLE{} and \EikeHE{} removes problematic event classes
and creates a sample with good data-to-simulation agreement before training the
boosted decision tree. At that point the normalization factor is $1.54$ and is
fixed for the rest of the analysis. The observed $\pm50\%$ variation of this
normalization factor at early cut levels is used as an estimate for the total
systematic uncertainty on the muon rate for samples \EikeLE{} and \Stephanie{},
which estimate the atmospheric muon background from simulation. The variation
encompasses uncertainties on the cosmic-ray spectrum and composition, the
particle interactions within air showers, the DOM efficiencies (see below) and
the optical properties of the ice. About $10\%$ of it can be attributed to the
uncertainty of the cosmic-ray energy spectrum. This value is obtained by varying
the broken power law parameters in the range of their published
uncertainties\cite{Glasstetter:1999ua}. Optimally, the other uncertainties would
be quantified by varying the respective parameters in the simulation. However,
the generation of several additional sufficiently sized background samples is
computationally intractable and we are hence left with the empirical estimate of
$50\%$ uncertainty for the atmospheric muon background. 

For sample \EikeHE{} the atmospheric muon background is estimated from a fit of
the reconstructed energy distribution and its extrapolation to higher energies.
From varying the fitted parameters within the uncertainties reported by the fit,
the systematic uncertainty of the muon rate in sample \EikeHE{} is estimated to
be $(-50\%,+150\%)$. This uncertainty is larger than for samples \EikeLE{} and
\Stephanie{}. However, for sample \EikeHE{} the background contribution of
atmospheric muons is much smaller than the contribution from atmospheric
neutrinos, which consequently dominate the total systematic uncertainty of the
combined background. 

A DOM's efficiency is the ratio of the light collected by a DOM to the total
light incident upon that DOM. The DOM efficiency includes the quantum
efficiency of the PMT and the transmissivity of the optical gel and glass of
each sphere. A $\pm10\%$ uncertainty in DOM efficiency is estimated for IceCube
DOMs\cite{Abbasi:2010vc}. By changing the DOM efficiency in the simulation the
effect on astrophysical (atmospheric) neutrinos event rates can be quantified to
$\pm4\%$ ($\pm14\%$) for sample \EikeLE{}\, and $\pm4\%$ ($\pm17\%$) for samples
\EikeHE{} and \Stephanie{}.

The simulation for this analysis assumed neutrino-nucleon cross sections based
on CTEQ5 parton distributions\cite{Gandhi:1995tf}.  The updated
CSS\cite{CooperSarkar:2007cv} calculation using the ZEUS global PDF fit
predicts smaller cross sections.  By comparing simulated neutrino datasets with
both models the systematic uncertainty from the neutrino cross-section model is
quantified as $\pm6\%$ ($\pm3\%$) for astrophysical (atmospheric) neutrinos.

The uncertainty in the atmospheric neutrino flux prediction has two components:
the theoretical uncertainty from the original calculations and the uncertainty
in modifying the HKKMS07 model to include the atmospheric neutrino knee. The
theoretical uncertainty of the conventional neutrino flux in the HKKMS07 model
is about $25\%$\cite{Honda:2006qj}. Since the ERS model is used as a baseline
for the prompt component its uncertainties are adopted\cite{Enberg:2008te}.
Combined, these result in a systematic uncertainty of the atmospheric neutrino
flux of $(-26\%,+25\%)$ and $(-37\%,27\%)$ in samples \EikeLE{} and \EikeHE{},
respectively. For sample \Stephanie{} we conservatively assume the uncertainty
from the high-energy sample \EikeHE{}. The knee in the cosmic-ray spectrum
should lead to a similar feature in the atmospheric neutrino spectrum. The
effect depends on the respective model for the cosmic-ray spectrum and the
energy transfer from the primary to the neutrino. Due to the different energy
ranges the samples are affected differently. The uncertainty was quantified to
be $(-16\%,+0\%)$ for sample \EikeLE{} and $(-23\%,+4\%)$ for sample \EikeHE{}.
For sample \Stephanie{} we adopt the larger values of the high-energy sample
\EikeHE{}.

Table \ref{Table:EikeSystematics} shows the resulting systematic uncertainty for
the various samples, where the total uncertainties are obtained from adding each
systematic uncertainty in quadrature. 


\section{Results \& Discussion} \label{Section:ResultsAndDiscussion}
We have prepared three different event selections of cascade-like events,
\EikeLE{}, \EikeHE{} and \Stephanie{}, each aiming at somewhat different
energies. In the following, we discuss the results starting with \EikeLE{}, the
analysis with the lowest energy threshold, and then move up in energy. The
reported results refer to $90\%$ of the experimental data, which were kept blind
during the development of the event selections. A summary of the results from the three
analyses is presented in Table \ref{Table:EikeUnblindingResults}.

\begingroup
\squeezetable
\begin{table}[tb]
\begin{ruledtabular}
\begin{tabular}{@{}lcccc@{}}
&\multicolumn{2}{c}{Sample \EikeLE{} (low energies)} & \multicolumn{2}{c}{Sample \EikeHE{} (high energies)} \\ \hline
                     & Atm. $\nu$     & $E^{-2} \nu$ & Atm. $\nu$     & $E^{-2} \nu$ \\ \hline      
DOM efficiency       & $14\%$         & $4\%$        & $17\%$         & $4\%$       \\       
Ice model            & $ 24\%$        & $11\%$       & $24\%$         & $11\%$      \\        
$\nu$ cross sections & $3\%$          & $6\%$        & $3\%$          & $6\%$       \\ 
Theoretical          & $-26\%\,+25\%$ & n/a          & $-37\%\,+27\%$ & n/a         \\        
Neutrino knee        & $-16\%$ $+0\%$ & n/a          & $-23\%$ $+4\%$ & n/a         \\ \\
Total                & $-41\%\,+37\%$ & $13\%$       & $-50\%\,+40\%$ & $13\%$      \\ \\
                     & \multicolumn{2}{c}{Atm. $\mu$ (simulated)} & \multicolumn{2}{c}{ Atm. $\mu$ (extrapolated)} \\ \hline
Total                &\multicolumn{2}{c}{$50\%$}     &\multicolumn{2}{c}{$-50\% +150\%$}   \\ 
\end{tabular}
\caption{Overview on systematics uncertainties on the event count for analyses
\EikeLE{} and \EikeHE{}. The systematic uncertainties associated with sample
\Stephanie{} are similar to those of sample \EikeHE{}. See section
\ref{Section:SystematicUncertainties} for further details.}
\label{Table:EikeSystematics}
\end{ruledtabular}
\end{table}
\endgroup

The sample \EikeLE{} with an energy threshold of about
$\unit{2}{\tera\electronvolt}$ aimed at the observation of atmospheric
neutrinos, which for this sample will be called the signal. In total $67$ events
were observed over an expectation of $41.1$ events from atmospheric muons and
$27.8$ from conventional and $2.25$ prompt atmospheric neutrinos, respectively.
Accordingly, the excess above atmospheric muons is quantitatively well described
by the atmospheric neutrino prediction by the HKKMS07 and ERS models. The rather
large uncertainty of the atmospheric muon background requires a careful
evaluation of the significance of the atmospheric neutrino excess. We
marginalize over the uncertainty in background and signal prediction using the
method described in Appendix \ref{Section:BayesianMethod}. The $90\%$ credible
interval ranges from $5$ to $62$ non-background events, or $16\%$ to $206\%$ of
the predicted conventional and prompt neutrino flux. The significance of the
excess over atmospheric muons including systematic errors is $1.1\sigma$. No
observation of atmospheric neutrinos is claimed. The average event energies of
atmospheric neutrinos are $\unit{6}{\tera\electronvolt}$ and hence comparable to
the highest energy bin of the completed analyses of contained events inside the
DeepCore/IceCube $79$-string configuration\cite{Aartsen:2012uu}. 

In sample \Stephanie{} with an energy threshold of about
$\unit{25}{\tera\electronvolt}$ we observed $14$ events after
event selection, on an expected background of $3.0$ atmospheric neutrino events
and $7.7$ atmospheric muon events. As the analyses was optimized for highest
sensitivity towards a diffuse $E^{-2}$ spectrum --- harder than the spectrum of
atmospheric neutrino and muon events --- the analysis has a higher energy
threshold. However, it was only realized after unblinding, that protons were
underrepresented at high energies in the simulation of the cosmic-ray spectrum
(see section \ref{Section:Simulation}), leading to an underrepresentation of the
muon background and suboptimal loose final cuts. As a result, the purity of the
sample is only comparable to that of sample \EikeLE{}. A small and insignificant
excess of events is observed over the background of atmospheric neutrinos and
muons. We calculate an all-flavor flux limit\cite{Hill:2002nv} using the TRolke
method\cite{Lundberg:2009iu} to include systematic errors. For an $E^{-2}$
astrophysical neutrino spectrum and assuming a $1\mcolon1\mcolon1$ flavor ratio
at the Earth, the all-flavor flux limit at a $90\%$ confidence level is
\begin{equation}
E^{2}\Phi_{\mathrm{lim}} \leq 7.46\times10^{-8} \flux.
\end{equation}
The energy range for this calculation containing $90\%$ of the signal is from
$\unit{25}{\tera\electronvolt}$ to $\unit{5}{\peta\electronvolt}$. 

\begin{figure}[tb]
\includegraphics[width=0.49\textwidth]{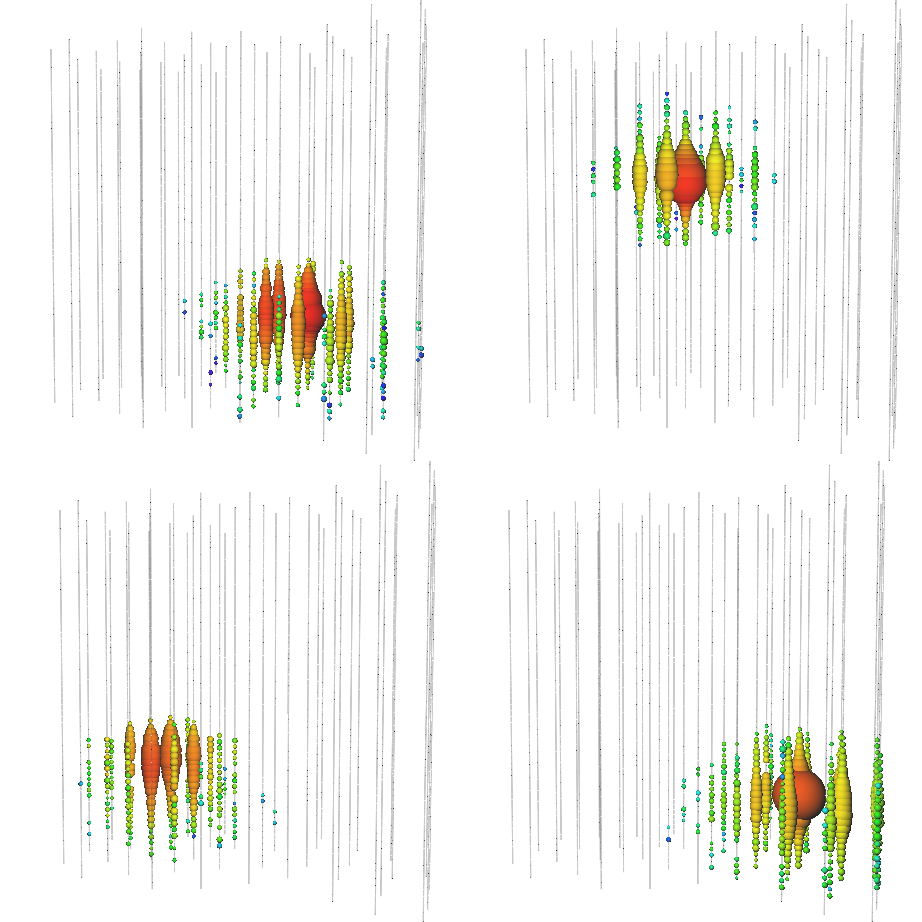}
\caption{Three cascade-like events above $\unit{100}{\tera\electronvolt}$ were
found in the samples \Stephanie{} and \EikeHE{}. An additional event was already
found in the $10\%$ sample used to develop the cuts. The event displays
visualize the light distribution in the detector: Each DOM is shown as a sphere,
which size scales with the recorded charge by the DOM. The color coding
illustrates the arrival time of the light at the DOM ranging from red (early
hits) over green to blue (late hits).}
\label{Figure:EikeEventDisplays}
\end{figure}

The sample \EikeHE{}, with an energy threshold of
$\unit{100}{\tera\electronvolt}$, is the high-energy counterpart of sample
\EikeLE{}, differing only in the lower BDT cut and higher energy cut. The cuts
were optimized for highest sensitivity towards an $E^{-2}$ spectrum and the
tighter cuts lead to a purer neutrino sample. In total $3$ events were found
over an expectation of $0.04$ from atmospheric muons and $0.21$ from atmospheric
conventional and prompt neutrinos. At these higher energies the expected
contribution from prompt neutrinos exceeds the conventional neutrinos by a
factor of $3$. One additional event with more than
$\unit{100}{\tera\electronvolt}$ was also in the $10\%$ sample used to develop
the analysis. It is not considered in the significance calculation. Images
illustrating the hit pattern of all four events are shown in
Fig.~\ref{Figure:EikeEventDisplays}.

\begin{figure}[tb]
\includegraphics[width=0.49\textwidth]{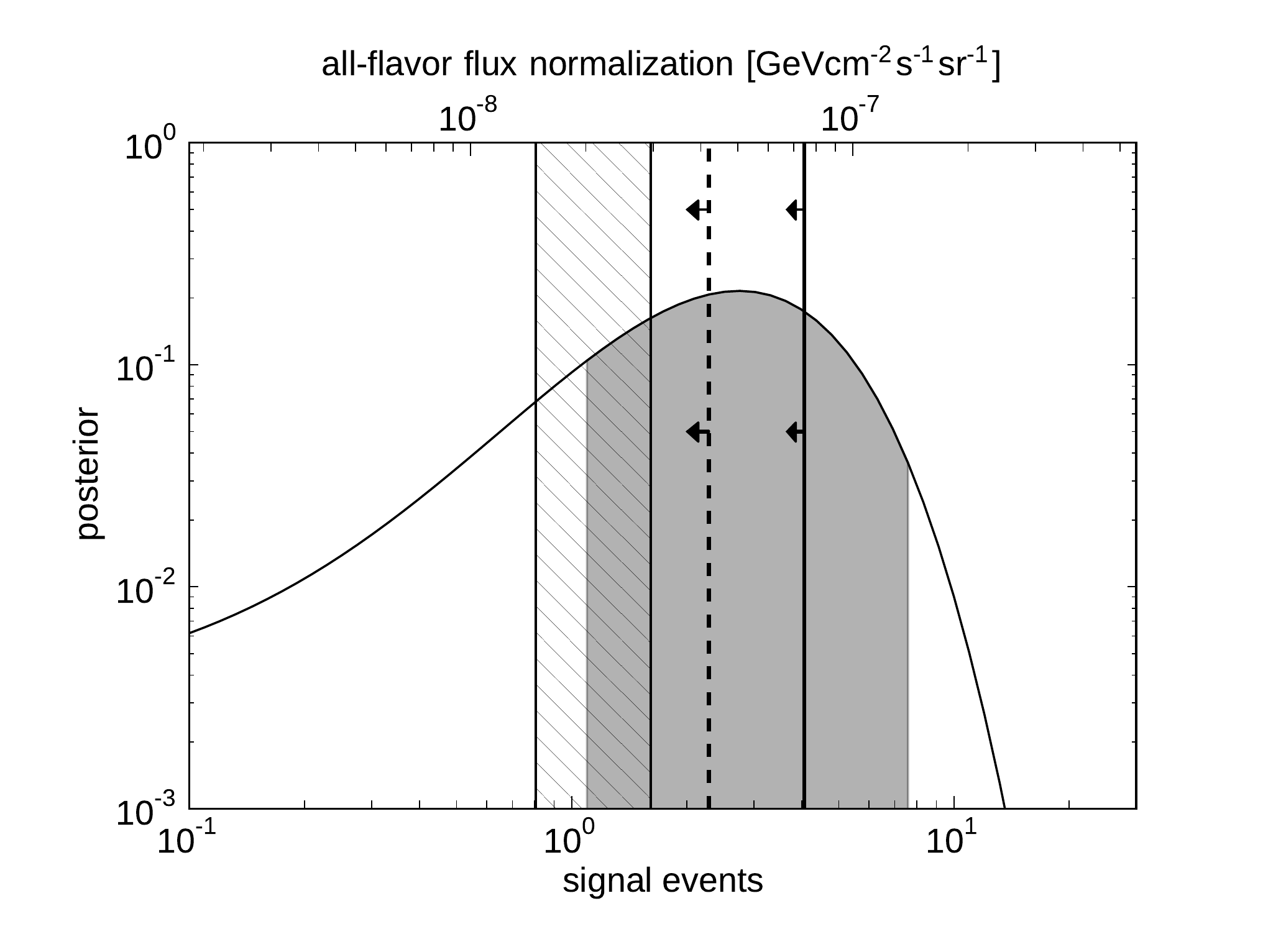}
\caption{ Posterior probability for additional signal events in the sample
\EikeHE{}. The gray shaded area shows the $90\%$ credible interval of this work
and the hatched area denotes the expected number of events according to the best
fit flux in\cite{hese}, including a cutoff at $\unit{2}{\peta\electronvolt}$.
The solid and dashed lines show the $90\%$ CL upper limits derived from sample
\Stephanie{} and the IceCube-$59$ diffuse muon search\cite{Schukraft:2013ya},
respectively. The upper $x$-axis transforms an expected event count into an
all-flavor flux normalization of an unbroken $E^{-2}$ power-law. In transforming
the flux estimate and cutoff from\cite{hese} to an unbroken power-law, a 
slightly lower flux normalization is obtained.}
\label{Figure:Posterior_highE}
\end{figure}

The $3$ events found are a rather large excess, not only above the muonic
background but also above the atmospheric neutrinos. It corresponds to
$2.7\sigma$ above both classes of background. We have employed the method
described in Appendix \ref{Section:BayesianMethod} allowing us the inclusion of
all systematic and statistical errors on the background expectation to calculate
the posterior probability for the potential signal flux shown in 
Fig.~\ref{Figure:Posterior_highE}. As a consequence of the observed excess the
posterior peaks around a flux normalization for an unbroken all-flavor $E^{-2}$
flux of $\unit{5\times10^{-8}}{\flux}$. The $90\%$ credible interval covers the
range $\unit{(2-14)\times 10^{-8}}{\flux}$. For a $1:1:1$ flavor ratio at Earth
of the expected events, $64\%$ would stem from electron neutrinos, $23\%$ from
tau neutrinos and $13\%$ from muon neutrinos.

The flux estimate derived from sample \EikeHE{} is compatible with the
astrophysical flux derived in\cite{hese}, taking into account the systematic
uncertainties.
The flux is higher than the upper limit found in a search for high-energy muon
neutrinos with the larger IceCube-$59$\cite{Schukraft:2013ya}. This is not
necessarily a contradiction, since the upper limit is set under the assumption
of an unbroken power law - a practical premise until measurements of high
energetic neutrinos provide a handle on any cutoff in the spectrum. A
non-equalized flavor ratio or a slightly different slope of the neutrino
spectrum could explain this, too.

\begingroup
\squeezetable
\begin{table}[t]
\begin{ruledtabular}
\begin{tabular}{@{}lccc@{}}
Sample                                                             & \EikeLE{}                      & \Stephanie{}                    & \EikeHE{}                        \\ \hline
$E_{\mathrm{cut}}$                                                 & $\unit{2}{\tera\electronvolt}$ & $\unit{25}{\tera\electronvolt}$ & $\unit{100}{\tera\electronvolt}$ \\ \hline
$10\%$ sample                                                      & $7$                            & $2$                             & $1$                              \\
$90\%$ sample                                                      & $67$                           & $12$                            & $3$                              \\ \hline
atm. $\mu$                                                         & $41.1 \pm 9.5$                 & $7.7 \pm 1.8$                   & $0.04$                           \\ \hline
Bartol (no knee)\cite{Barr:2004br}                                 & $25.5 \pm 2.8$                 & $2.12 \pm 0.2$                  & $0.078 \pm 0.012$                \\
HKKMS07 (+knee)\cite{Honda:2006qj,Gaisser:2012zz,Schukraft:2013ya} & $27.8 \pm 3.0$                 & $1.68 \pm 0.16$                 & $0.054 \pm 0.009$                \\ \hline
ERS\cite{Enberg:2008te} (max.)                                     & $2.76 \pm 0.07$                & $1.21 \pm 0.03$                 & $0.198 \pm 0.005$                \\
ERS\cite{Enberg:2008te}                                            & $2.25 \pm 0.06$                & $0.95 \pm 0.02$                 & $0.155 \pm 0.004$                \\
ERS\cite{Enberg:2008te} (min.)                                     &  $1.29 \pm 0.03$               & $0.57 \pm 0.01$                 & $0.090 \pm 0.002$                \\
Martin GBW\cite{Martin:2003us}                                     & $1.14 \pm 0.03$                & $0.48 \pm 0.01$                 & $0.078 \pm 0.002$                \\
$3.6\times10^{-8} E^{-2}$                                          & $4.54 \pm 0.12$                & $4.93 \pm 0.12$                 & $1.96 \pm 0.05$                  \\

\end{tabular}
\caption{Event count predictions and results for the different samples and for
models of conventional, prompt and astrophysical neutrinos. The model
predictions are calculated for $90\%$ of the experimental data. Where they are
derived from simulation the statistical errors are given. The three lines of the
ERS model show the uncertainty band.}
\label{Table:EikeUnblindingResults}
\end{ruledtabular}
\end{table}
\endgroup

The results from the sample \EikeLE{}, \EikeHE{} and \Stephanie{} are
consistent, with an excess appearing only towards larger energies. In addition,
one can compare the overlap between the samples as a crosscheck. Due to the
different energy thresholds, one expects sample \Stephanie{} to be largely a
subsample of \EikeLE{}. However, due to the difference in selection criteria and
general low efficiency of only $\sim10\%$ with which signal events are selected,
a perfect overlap can not be expected. We find that of the $14$ events of sample
\Stephanie{}, $7$ are also contained in \EikeLE{}, where the majority of the
remaining have not passed the different containment cut. All three high-energy
events of sample \EikeHE{} are also contained in sample \Stephanie{}. Only two
of them are also in sample \EikeLE{}.


\section{Summary} \label{Section:Summary}
We have reported on a search for astrophysical and atmospheric neutrino-induced
particle showers in the IceCube-$40$ detector. The data was taken between April
2008 and May 2009 with a total of $367.1$ days livetime.

Three different event selections, \EikeLE{}, \EikeHE{} and \Stephanie{}, each
tailored to different energy ranges, were developed and applied to the data. The
low and intermediate energy samples \EikeLE{} and \Stephanie{} have an energy
threshold of $\unit{2}{\tera\electronvolt}$ and $\unit{25}{\tera\electronvolt}$,
respectively, and have sensitivity to atmospheric neutrinos. In both samples, an
excess over the atmospheric muon background was observed that can be well
explained by conventional atmospheric neutrinos.

With an energy threshold of $\unit{25}{\tera\electronvolt}$, analysis
\Stephanie{} also has sensitivity to a high-energy astrophysical neutrino flux.
In the absence of an excess of events above the combined background of
atmospheric neutrinos and muons we are left with an upper limit ($90\%$
confidence level) of $\unit{E^2 \Phi_{lim}\leq7.46\times10^{-8}}{\flux}$ on the
all-flavor astrophysical neutrino flux, assuming an $E^{-2}$ spectrum and a
$1\mcolon1\mcolon1$ neutrino flavor ratio at the Earth. For this limit, $90\%$
of the expected signal events have energies in the range between
$\unit{25}{\tera\electronvolt}$ to $\unit{5}{\peta\electronvolt}$. The upper
limit is below that reported from cascade searches using the IceCube-$22$
detector\cite{Abbasi:2011ui} and approaches the Waxman-Bahcall
limit\cite{Waxman:1998yy,Waxman:2011hr}.

Finally, sample \EikeHE{} was optimized towards the largest sensitivity for high
energy astrophysical neutrinos. Compared to sample \Stephanie{}, it profited
from a larger sample of simulated muon background during the optimization of the
event selection. In $90\%$ of the available data, three events were observed
above the energy threshold of $\unit{100}{\tera\electronvolt}$, with an
expectation of only $0.25$ events from atmospheric neutrinos (both conventional
and prompt) as well as atmospheric muons --- a $\unit{2.7}{\sigma}$ excess. The
$10\%$ burn sample contains a fourth event, which has not entered the
significance calculation.

The excess seen above $\unit{100}{\tera\electronvolt}$ in sample \EikeHE{} is
noteworthy. Although not significant enough to claim evidence for an
astrophysical neutrino flux, it is consistent with similar excesses found in
diffuse neutrino searches with IceCube-$59$\cite{Schukraft:2013ya} using muon
neutrino events, the two $\unit{}{\peta\electronvolt}$
events\cite{Aartsen:2013bka} and the $28$ events found in\cite{hese}, the
strongest single evidence that IceCube is seeing a high-energy neutrino flux of
astrophysical neutrinos. This analysis thereby provides three neutrino event
candidates between $\unit{140}{\tera\electronvolt}$ and
$\unit{220}{\tera\electronvolt}$ --- an intermediate energy scale --- with
unprecedented low background contamination of conventional atmospheric
neutrinos and muons. The constraints on the all-flavor normalization of the
high-energy neutrino flux of astrophysical neutrinos with equal flavor
contributions are summarized in Fig.~\ref{Figure:Posterior_highE}.  A $90\%$
credible interval covers the range $\unit{(2-14)\times10^{-8}}{\flux}$ and is
compatible with the more stringent flux estimate established by\cite{hese}.

However, the sensitivity towards a diffuse flux of high-energy neutrinos was
reached using data from only $50\%$ of the final IceCube configuration. It is
the good energy resolution and the little intrinsic background associated with
the signature of neutrino-induced cascades, that is thereby providing the large
sensitivity to the diffuse flux\cite{Kowalski:2005tz}. The IceCube detector is
now completed with $86$ strings and an instrumented volume of
$\unit{1}{\kilo\metre\cubed}$. Future cascade searches benefit from the more
favorable detector geometry, which allows for even better suppression of the
background from atmospheric muons and improved cascade detection
efficiency\cite{Schoenwald:2013ab,Bzdak:2013ab}.
Neutrino-induced cascades will hence continue to play a prime role in further
exploring the high-energy astrophysical neutrino flux.


\begin{appendix}

\section{Calculation of the posterior probabilities} \label{Section:BayesianMethod}
The experiment under consideration is a counting experiment in the presence
of background. In order to incorporate systematic and statistical uncertainties 
into the interpretation of the result, a Bayesian approach was chosen. The
probability to have observed $n_{\mathrm{obs}}$ events in the presence of
$n_{\mathrm{sig}}$ events, signal selection efficiency $\epsilon$ and
$n_{\mathrm{bg}}$ background events is given by the Poisson probability:
\begin{equation}
P(n_{\mathrm{obs}}|n_{\mathrm{sig}},n_{\mathrm{bg}}) = \frac{(\epsilon n_{\mathrm{sig}}+n_{\mathrm{bg}})^{n_{\mathrm{obs}}}}{n_{\mathrm{obs}}!}\exp\left(-(\epsilon n_{\mathrm{sig}}+n_{\mathrm{bg}})\right) .
\end{equation}
From having observed $n_{\mathrm{obs}}$ we want to infer, whether a given value
of $n_{\mathrm{sig}}$ is supported or ruled out by the experimental result.
This information is given by the posterior probability of $n_{\mathrm{sig}}$,
which can be obtained by applying Bayes theorem. In order to use the theorem,
available information on the expected signal as well as the uncertainty of the
other parameters must be quantified in the form of priors. By marginalizing over
all parameters other than $n_{\mathrm{sig}}$, remaining uncertainties are then
incorporated into the final result.

The background uncertainties from model predictions and selection efficiencies
are described by $p(n_{\mathrm{bg}})$. For the signal we model all uncertainties
with $p(\epsilon)$ and the prior belief with $p(n_{\mathrm{sig}})$.
The posterior probability can then be calculated:
\begin{equation}
P(n_{\mathrm{sig}}|n_{\mathrm{obs}}) \propto \int dn_{\mathrm{bg}}d\epsilon P(n_{\mathrm{obs}}|n_{\mathrm{sig}},n_{\mathrm{bg}}) p(n_{\mathrm{bg}}) p(\epsilon) p(n_{\mathrm{sig}})
\end{equation}

A constant is chosen for the signal prior to reflect no prior knowledge on the
signal. For the background prior a Gaussian is used. The mean is centered at
the rate prediction, the width represents the modeled uncertainty and it is
truncated at zero since rates have to be positive. The uncertainty in the signal
efficiency is modeled with the factor $\epsilon$ that is applied to the number
of signal events after all cuts. The prior for $\epsilon$ is modeled with
a Gaussian centered at $1$ and having a width corresponding to the systematic
uncertainty. It is also truncated at zero.
\end{appendix}

\begin{acknowledgments}
We acknowledge the support from the following agencies:
U.S. National Science Foundation-Office of Polar Programs,
U.S. National Science Foundation-Physics Division,
University of Wisconsin Alumni Research Foundation,
the Grid Laboratory Of Wisconsin (GLOW) grid infrastructure at the University of Wisconsin - Madison, the Open Science Grid (OSG) grid infrastructure;
U.S. Department of Energy, and National Energy Research Scientific Computing Center,
the Louisiana Optical Network Initiative (LONI) grid computing resources;
Natural Sciences and Engineering Research Council of Canada,
WestGrid and Compute/Calcul Canada;
Swedish Research Council,
Swedish Polar Research Secretariat,
Swedish National Infrastructure for Computing (SNIC),
and Knut and Alice Wallenberg Foundation, Sweden;
German Ministry for Education and Research (BMBF),
Deutsche Forschungsgemeinschaft (DFG),
Helmholtz Alliance for Astroparticle Physics (HAP),
Research Department of Plasmas with Complex Interactions (Bochum), Germany;
Fund for Scientific Research (FNRS-FWO),
FWO Odysseus programme,
Flanders Institute to encourage scientific and technological research in industry (IWT),
Belgian Federal Science Policy Office (Belspo);
University of Oxford, United Kingdom;
Marsden Fund, New Zealand;
Australian Research Council;
Japan Society for Promotion of Science (JSPS);
the Swiss National Science Foundation (SNSF), Switzerland;
National Research Foundation of Korea (NRF);
Danish National Research Foundation, Denmark (DNRF)
\end{acknowledgments}

\nocite{Abbasi:2009nfa}
\nocite{GonzalezGarcia:2006ay}

\bibliographystyle{unsrt}              

\end{document}